\documentclass{jfp-arxiv}

\usepackage[utf8]{inputenc}
\usepackage[T1]{fontenc}
\usepackage{listings,multicol}
\usepackage{elm-highlighting}
\usepackage{lstcoq}
\usepackage{listings-rust}
\usepackage{amsmath,amsthm}
\usepackage{amssymb}
\usepackage{subcaption}
\usepackage{semantic}
\usepackage[toc]{appendix}
\usepackage[inline]{enumitem}
\usepackage{hyperref}
\usepackage{booktabs}
\usepackage{pifont}
\usepackage{natbib}
\usepackage{graphicx}

\usepackage{xcolor}

\definecolor{ltblue}{rgb}{0,0.4,0.4}
\definecolor{dkblue}{rgb}{0,0.1,0.6}
\definecolor{dkgreen}{rgb}{0,0.5,0}
\definecolor{brightmaroon}{rgb}{0.76, 0.13, 0.28}
\definecolor{burntorange}{rgb}{0.8, 0.33, 0.0}
\definecolor{dkred}{rgb}{0.5,0,0}
\definecolor{bggray}{gray}{0.95}
\definecolor{arsenic}{rgb}{0.23, 0.27, 0.29}

\newcommand{\coqLogo}{\includegraphics[width=1.2em]{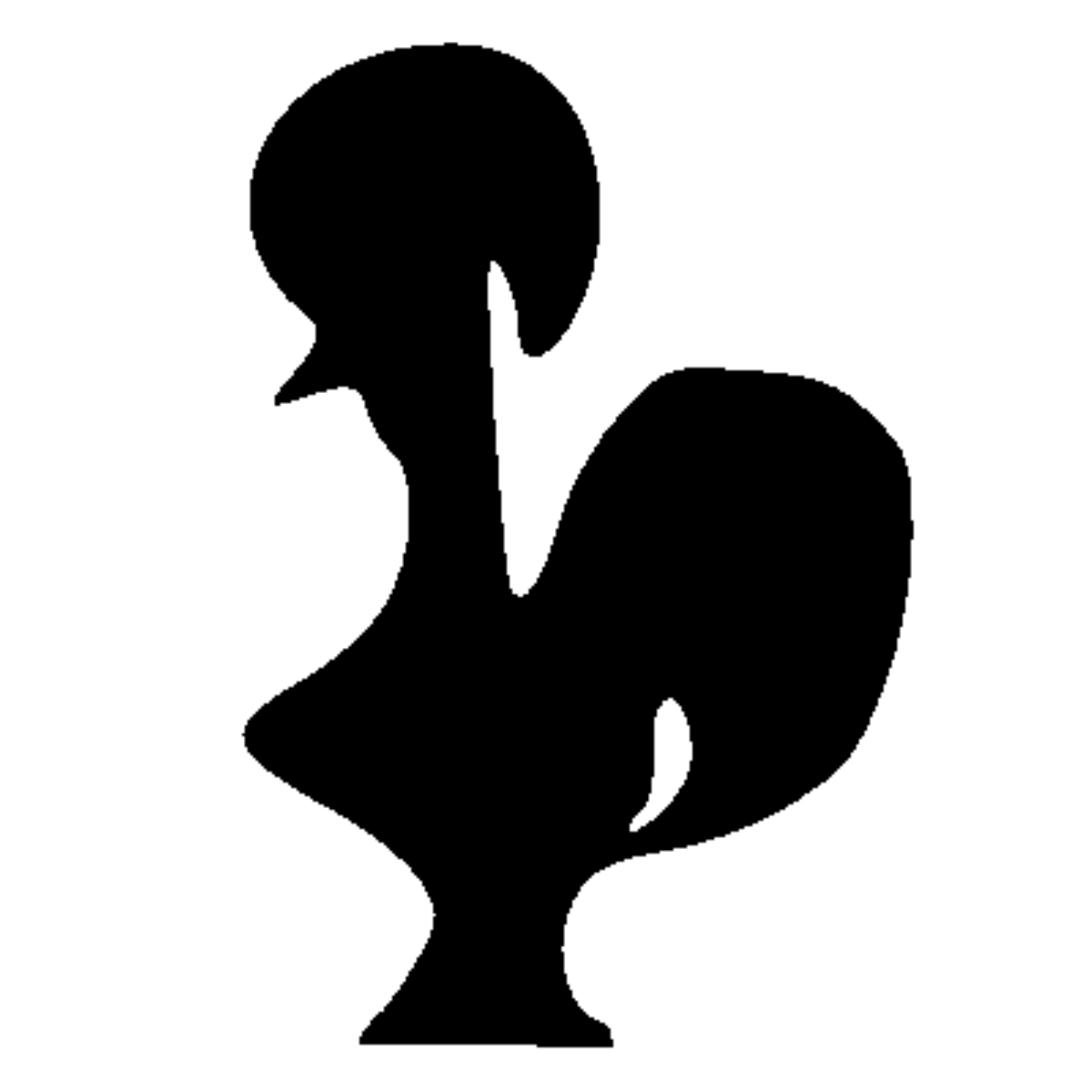}}

\newcommand{\icode}[1]{\lstinline[mathescape,basicstyle=\small]!#1!}
\makeatletter
\newtheoremstyle{coqtheorem}% name
  {}%         Space above, empty = `usual value'
  {}%         Space below
  {\itshape}% Body font
  {}%         Indent amount (empty = no indent, \parindent = para indent)
  {\bfseries}% Thm head font
  {}%        Punctuation after thm head
  {\newline}% Space after thm head: \newline = linebreak
  {\thmname{#1}\thmnumber{\@ifnotempty{#1}{ }#2}%
   \thmnote{ {\the\thm@notefont(#3).}\newline\extraref}}

\theoremstyle{coqtheorem}
\newtheorem{coqtheoreminner}{Theorem}
\newcommand{\extraref}{}
\newenvironment{coqtheorem}[1]
 {%
  \if\relax\detokenize{#1}\relax
  \else
    \renewcommand{\extraref}{#1}% if argument is empty, add nothing
  \fi
  \begin{coqtheoreminner}}
 {\end{coqtheoreminner}}

\theoremstyle{definition}

\newtheoremstyle{coqdefinition}% name
  {}%         Space above, empty = `usual value'
  {}%         Space below
  {\normalfont}% Body font
  {}%         Indent amount (empty = no indent, \parindent = para indent)
  {\bfseries}% Thm head font
  {}%        Punctuation after thm head
  {\newline}% Space after thm head: \newline = linebreak
  {\thmname{#1}\thmnumber{\@ifnotempty{#1}{ }#2}%
   \thmnote{ {\the\thm@notefont(#3).}\newline\extrarefdef}}

\theoremstyle{coqdefinition}
\newtheorem{coqdefinitioninner}{Definition}
\newcommand{\extrarefdef}{}
\newenvironment{coqdefinition}[1]
 {%
  \if\relax\detokenize{#1}\relax
  \else
    \renewcommand{\extrarefdef}{#1}% if argument is empty, add nothing
  \fi
  \begin{coqdefinitioninner}}
 {\end{coqdefinitioninner}}

\newcommand\CIC[0]{CIC}
\newcommand\CICbox[0]{\ensuremath{\lambda_\square}}
\newcommand\IR[0]{\ensuremath{\lambda^T_\square}}

\newcommand\lift[1]{\uparrow_1#1}

\reservestyle{\command}{\mathtt}
\command{if[if],then[then],else[else],let[let],in[in],case[case],return[return],of[of],fix[fix],
fun[fun],Set[Set],forall[forall],match[match],as[as],in[in],with[with],end[end],
NotPrenex[NotPrenex],Ok[Ok],TypeError[TypeError],TypingError[TypingError],
Error[Error],NotSupported[NotSupported],Checked[Checked],true[true],false[false],
ret[ret],isarity[is\_arity],islogical[is\_logical],issort[is\_sort],convtoarity[conv\_ar],
tSort[Type],RelOther[Other],RelTypeVar[TV],RelInductive[Ind],
tConst[C],tInd[I],Some[Some],None[None]}

\colorlet{cicterms}{dkblue}
\colorlet{boxtyterms}{dkgreen}
\colorlet{errorcolor}{red}

\newcommand\cic[1]{{{\color{cicterms}{#1}}}}

\newcommand\boxty[1]{{\color{boxtyterms}{#1}}}
\newcommand\TBox{\boxty{\square}}
\newcommand\TAny{\boxty{\ensuremath{\mathbb{T}}}}
\newcommand\flagoftype{{\mathtt{flag\_of\_type}}}

\newcommand\ident[1]{{\mathit{#1}}}
\newcommand\erasetypename{\ensuremath{\mathcal{E}^T}}
\newcommand\erasetype[4]{\erasetypename{}~#1~#2~#3~#4~}
\newcommand\erasetypeappname{\mathcal{E}_{app}^T}
\newcommand\erasetypeapp[4]{\erasetypeappname{}~#1~#2~#3~#4~}
\newcommand\erasetypeheadname{{\ensuremath{\mathcal{E}_{head}^T}}}
\newcommand\erasetypehead[2]{\erasetypeheadname{}~#1~#2}
\newcommand\erasevarname{{\ensuremath{\mathcal{E}_{var}^T}}}
\newcommand\erasevar[2]{{\erasevarname{}~#1~#2}}
\newcommand\erasetypeschemename{\ensuremath{\mathcal{E}^{TS}}}
\newcommand\erasetypescheme[5]{\erasetypeschemename{}~#1~#2~#3~#4~#5~}
\newcommand\erasetypeschemeetaname{\ensuremath{\mathcal{E}_\eta^{TS}}}
\newcommand\erasetypeschemeeta[5]{\erasetypeschemeetaname{}~#1~#2~#3~#4~#5~}

\newcommand\type[1]{\ensuremath\mathtt{#1}}

\newcommand\dearg{{\ident{dearg}}}
\newcommand\deargcst{{\ident{dearg\_cst}}}
\newcommand\deargmib{{\ident{dearg\_mib}}}
\newcommand\deargenv{{\ident{dearg\_env}}}

\newcommand\wcbveval[3]{{\ensuremath{#1 \vdash #2 \triangleright #3}}}
\newcommand\wcbvevalpcuic[3]{{\ensuremath{#1 \vdash_p #2 \triangleright #3}}}

\newcommand\redbetaiotazeta{{\mathtt{red}_{\beta\iota\zeta}}}
\newcommand\decomposeapp{{\mathtt{decompose\_app}}}
\newcommand\incvar{{\mathtt{inc\_var}}}
\newcommand\isnone{{\mathtt{is\_none}}}
\newcommand\infer{{\mathtt{infer}}}
\newcommand\fst{{\mathtt{fst}}}
\newcommand\snd{{\mathtt{snd}}}
\newcommand\canhaveargs{{\mathtt{can\_have\_args}}}

\newcommand\coqref[2]{\raisebox{-0.3em}{\protect\coqLogo}{\footnotesize\href{https://github.com/AU-COBRA/ConCert/blob/#2}{\texttt{\color{brightmaroon}{#1}}}}}

\begin{document}

%% \journaltitle{JPF}
%% \cpr{Cambridge University Press}
%% \doival{}

%% \totalpg{\pageref{lastpage01}}
%% \jnlDoiYr{2021}

\title{Extracting functional programs from Coq, in Coq}

\begin{authgrp}
  \author{Danil Annenkov}
  \affiliation{Concordium Blockchain Research Center, Aarhus University}
  \author{Mikkel Milo}
  \affiliation{Department of Computer Science, Aarhus University, Denmark}
  \author{Jakob Botsch Nielsen}
  \affiliation{Concordium Blockchain Research Center, Aarhus University}
  \author{Bas Spitters}
  \affiliation{Concordium Blockchain Research Center, Aarhus University}
\end{authgrp}

\lefttitle{Extracting functional programs from Coq, in Coq}
\righttitle{Annenkov, Milo, Nielsen, Spitters}

\begin{abstract}
  We implement extraction of Coq programs to functional languages based on MetaCoq's certified erasure.
  We extend the MetaCoq erasure output language with typing information and use it as an intermediate representation, which we call $\IR$.
  We complement the extraction functionality with a full pipeline that includes several standard transformations (eta-expansion, inlining, etc) implemented in a proof-generating manner along with a verified optimisation pass removing unused arguments.
  We prove the pass correct wrt.\ a conventional call-by-value operational semantics of functional languages.
  From the optimised $\IR$ representation, we obtain code in two functional smart contract languages, Liquidity and CameLIGO, the functional language Elm, and a subset of the multi-paradigm language for systems programming Rust.
  Rust is currently gaining popularity as a language for smart contracts, and we demonstrate how our extraction can be used to extract smart contract code for the Concordium network.
  The development is done in the context of the ConCert framework that enables smart contract verification.
  We contribute with two verified real-world smart contracts (boardroom voting and escrow), which we use, among other examples, to exemplify the applicability of the pipeline.
  In addition, we develop a verified web application and extract it to fully functional Elm code.
  In total, this gives us a way to write dependently typed programs in Coq, verify, and then extract them to several target languages while retaining a small trusted computing base of only MetaCoq and the pretty-printers into these languages.
\end{abstract}

\maketitle

\lstset{language = Coq}
\lstset{basewidth = 0.5em}

\section{Introduction}\label{sec:intro}
Proof assistants offer a promising way of delivering the strongest guarantee of correctness.
Many software properties can be stated and verified using the currently available tools such as e.g.\ Coq, Agda, Isabelle.
In the current work we focus our attention on the Coq proof assistant, based on dependent type theory (calculus of inductive constructions --- CIC).
Since the calculus of Coq is also a programming language, it is possible to execute programs directly in the proof assistant.
The expressiveness of Coq's type system allows for writing specifications directly in a program type.
These specification can be expressed, for example, in the form of pre- and postconditions using \emph{subset types} implemented in Coq using the dependent pair type ($\Sigma$-type).
However, in order to integrate the formally verified code with existing components, one would like to obtain a program in other programming languages.
One way of achieving this is to \emph{extract} the executable code from the formalised development.
Various verified developments rely extensively on the extraction feature of proof assistants~\citep{2006-Leroy-compcert,klein14tocs,ExtractionLargeProofs:CrusFilipeSpitters,ExecutingExtracted:CruzFilipeLetouzey,FillaitreLetouzey:FSets}.
However, currently, the standard extraction feature in proof assistants focuses on producing code in conventional functional languages (Haskell, OCaml, Standard ML, Scheme, etc.).
Nowadays, there are many new important target languages that are not covered by the standard extraction functionality.

An example of a domain that experiences rapid development and the increased importance of verification is the \emph{smart contract technology}.
Smart contracts are programs deployed on top of a blockchain.
They often control large amounts of value and cannot be changed after deployment.
Unfortunately, many vulnerabilities have been discovered in smart contracts and this has led to huge financial losses (e.g.\ TheDAO\footnote{\url{https://www.wired.com/2016/06/50-million-hack-just-showed-dao-human/}. Accessed: 2021-07-20}, Parity's multi-signature wallet\footnote{\url{https://www.parity.io/the-multi-sig-hack-a-postmortem/}. Accessed: 2021-07-20}).
Therefore, smart contract verification is crucially important.
Functional smart contract languages are becoming increasingly popular: e.g.\ Simplicity~\citep{O'Connor:Simplicity}, Liquidity~\citep{Liquidity}, Plutus~\citep{Chapman:PlutusCore}, Scilla~\citep{Sergey:ScillaOOPSLA} and the LIGO family\footnote{\url{https://ligolang.org/}. Accessed: 2021-07-20}.
A contract in such a language is a partial function from a message type and a current state to a new state and a list of actions (transfers, calls to other contracts), making smart contracts more amenable for formal verification.
Functional smart contract languages, similarly to conventional functional languages, are often based on a well-established theoretical foundation (variants of the Hindley-Milner type system).
The expressive type system, immutability and message-passing execution model allow for ruling out many common errors in comparison with conventional smart contract languages such as Solidity.

For the errors that are not caught by the type checker, a proof assistant, in particular Coq, can be used to ensure correctness.
Once properties of contracts are verified, one would like to execute them on blockchains.
At this point, the code extraction feature of Coq would be a great asset, but extraction to smart contract languages is not available in Coq.

There are other programming languages of interest in different domains that are not covered by the current Coq extraction.
Among these, Elm~\citep{ElmInAction} --- a functional language for web development and Rust~\citep{TheRustBook} --- a multi-paradigm systems programming language, are two examples.

Another issue we face is that the current implementation of Coq extraction is written in OCaml and is not itself verified, potentially breaking the guarantees provided by the formalised development.
We address this issue by using an existing formalisation of the meta-theory of Coq and provide a framework that is implemented Coq itself.
Being written in Coq gives us a significant advantage since it makes it possible to apply various techniques to verify the development itself.

\begin{figure}
  \centering
  \includegraphics[width=12.8cm]{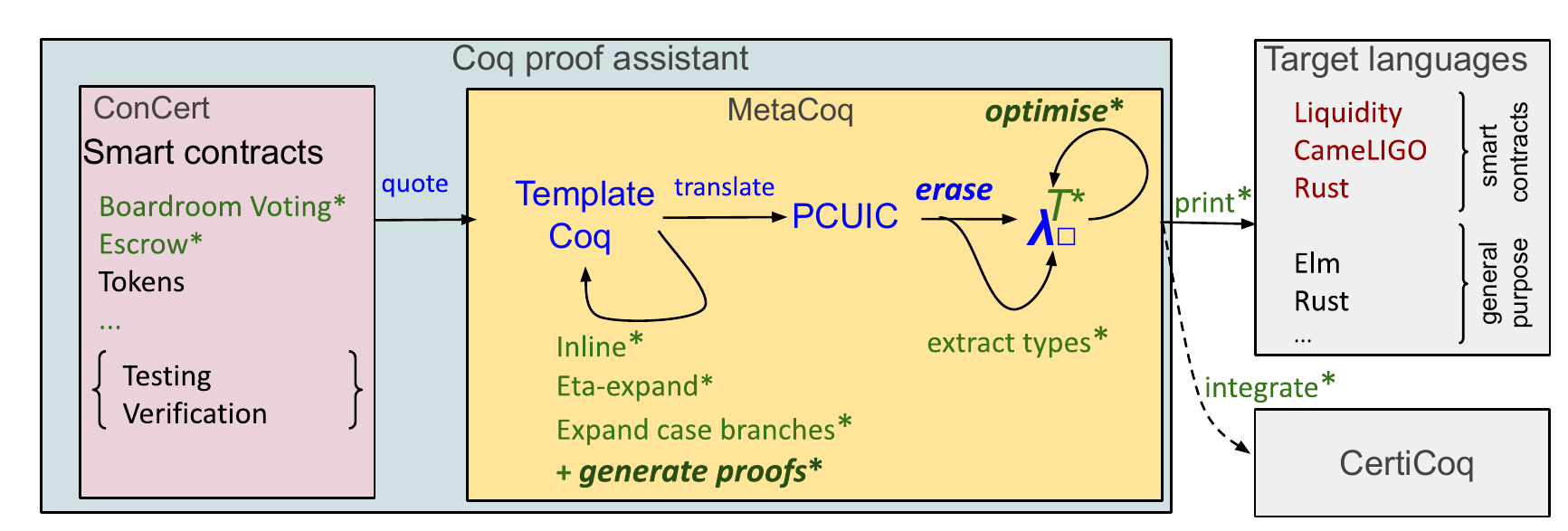}
  \caption{The pipeline}\label{fig:pipeline}
\end{figure}

The current work extends and improves the results previously published and presented by the same authors at the conference Certified Programs and Proofs~\citep{ConCert-extraction-testing} in January 2021.
We build on the ConCert framework~\citep{ConCert,Interactions} for smart contracts verification in Coq and the MetaCoq project~\citep{MetaCoq}.
We summarise the contributions as the following, marking with $^{\dagger}$ the contributions that extend the previous work.
\begin{itemize}
 \item We provide a general framework for extraction from Coq to a typed functional language (\cref{sec:extraction}).
   The framework is based on certified erasure~\citep{CertErasure} of MetaCoq.
   The output of MetaCoq's erasure procedure is an AST of an untyped functional programming language $\CICbox$.
   In order to generate code in typed programming languages, we implement an erasure procedure for types and inductive definitions.
   We add the typing information for all $\CICbox$ definitions and implement an annotation mechanism allowing for adding annotations in a modular way --- without changing the AST definition.
   We call the resulting representation $\IR$ and use it as an intermediate representation.
   Moreover, we implement and prove correct an optimisation procedure that removes unused arguments.
   The procedure allows us to optimise away some computationally irrelevant bits left after erasure.
 \item We implement pre-processing passes before the erasure stage (see~\cref{sec:proof-generating}). After running all the passes we generate correctness proofs.
   The passes include:
   \begin{itemize}
   \item $\eta$-expansion;
   \item expansion of \icode{match} branches$^{\dagger}$;
   \item inlining$^{\dagger}$.
   \end{itemize}
 \item We develop in Coq pretty-printers for obtaning extracted code from our intermediate representation to the following target languages.
   \begin{itemize}
     \item Liquidity --- a functional smart contract language for the Dune network (see~\cref{sec:liquidity-cameligo-extraction}).
     \item CameLIGO --- a functional smart contract language from the LIGO family for the Tezos network (see~\cref{sec:liquidity-cameligo-extraction})$^{\dagger}$.
     \item Elm --- a general purpose functional language used for web development (see~\cref{sec:elm-extraction}).
     \item Rust --- a multi-paradigm systems programming languages (see~\cref{sec:rust-extraction}).$^{\dagger}$.
   \end{itemize}
 \item We develop an integration infrastructure, required to deploy smart contracts written in Rust on the Concordium blockchain$^{\dagger}$.
 \item We provide case studies of smart contracts in ConCert by proving properties of an escrow contract and an anonymous voting contract based on the Open Vote Network protocol (\cref{sec:escrow,sec:boardroom}).
   We apply our extraction functionality to study the applicability of our pipeline to the developed contracts.
\end{itemize}

Apart from the extensions marked above, we have improved over the previous work in the following points.
\begin{itemize}
  \item The erasure procedure for types now covers type schemes.
    We provide the updated procedure along with the discussion in~\cref{sec:erasure-for-types}
  \item We extract the escrow contract to new target languages and finalise the extraction of the boardroom voting contract, which was not previously extracted.
    For the Elm extraction, we develop a verified web application that uses dependent types to encode the validity of the data in the application model.
    We demonstrate how the fully functional web application can be produced from the formalisation.
\end{itemize}

\section{The pipeline}
We begin by describing the whole pipeline covering the full process of starting with a program in Coq and ending with extracted code in one of the target languages.
This pipeline is shown in \cref{fig:pipeline}.
The items marked with $^{*}$ (also given in \boxty{\textsf{green}}) are contributions of this work and the items in \emph{\textbf{\textsf{bold cursive}}} are verified.
The MetaCoq project~\citep{MetaCoq} provides us with metaprogramming facilities (e.g.\ quoting Coq terms) and formalisation of the meta-theory of Coq, inlcuding the verified erasure procedure.

We start by developing a program in Gallina that can use rich Coq types in the style of certified programming (see e.g.\ \citep{Chlipala:cpdt}).
In the case of smart contracts, we can use the machinery available in ConCert to test and verify the properties of interacting smart contracts.
We obtain a Template Coq representation by quoting the program.
This representation is close to the actual AST representation in the Coq kernel.
We then apply a number of \emph{certifying} transformations to this representation (see \cref{sec:proof-generating}).
This means that we produce a transformed term along with a proof term, witnessing that the transformed term is equal to the original in the theory of Coq.
Currently, we assume that the transformations applied to the Template Coq representations preserve convertibility.
Therefore, we can easily certify them by generating simple proofs consisting essentially of the constructor of Coq's equality type \icode{eq_refl}.
Although the transformations themselves are not verified, generated proofs give strong guarantees that the behaviour of the term has not been altered.
One can configure the pipeline to apply several transformations, in this case, they will be composed together and applied to the Template Coq term.
The correctness proofs are generated after all the specified transformations are applied.

The theory of Coq is presented by the predicative calculus of cumulative inductive constructions (PCUIC)~\citep{TimanySozeau:PCUIC}, which is essentially a cleaned-up version of the kernel representation.
The translation from the Template Coq representation to PCUIC is mostly straightforward.
Currently, MetaCoq provides the type soundness proof for the translation, but computational soundness (wrt.\ weak call-by-value evaluation) is not verified.
However, the MetaCoq developers plan to close this gap in the near future.
Most of the meta-theoretic results formalised by the MetaCoq project use the PCUIC representation (see~\cref{sec:MetaCoq} for the details about different MetaCoq representations).

From PCUIC, we obtain a term in an untyped calculus of erased programs $\CICbox$ using the verified erasure procedure of MetaCoq.
By $\IR$, we denote $\CICbox$ enriched with typing information, which we obtain using our erasure procedure for types (see~\cref{sec:erasure-for-types}).
Specifically, we add to the $\CICbox$ representation of MetaCoq the following.
\begin{itemize}
\item Constants and definitions of inductive types in the global environment store the corresponding ``erased'' types ($\boxty{\type{box\_type}}$ in~\cref{sec:erasure-for-types}).
\item We explicitly represent \emph{type aliases} (definitions that accept some parameters and return a type) as entries in the extended global environment.
\item The nodes of the $\CICbox$ AST can be optionally annotated with the corresponding ``erased'' types (see~\cref{sec:liquidity-cameligo-extraction}).
\end{itemize}

The typing information is required for extracting to typed functional languages.
The $\IR$ representation is essentially a core of a pure statically typed functional programming language.
Our extensions make it a convenient intermediate representation containing enough information to generate code in various target languages.

The pipeline provides a way of specifying optimisations in a compositional way.
These optimisations are applied to the $\IR$ representation.
Each optimisation should be accompanied with a proof of computational soundness wrt.\ the big-step call-by-value evaluation relation for $\CICbox$ terms.
The format for the computational soundness is fixed and the individual proofs are combined in the top-level statement covering given optimisation steps (see~\cref{thm:extract-sound}).
At the current stage, we provide an optimisation that removes dead arguments of functions and constructors (see \cref{sec:optimisations}).

The optimised $\IR$ code is then can be printed using the pretty-printers developed directly in Coq.
The target languages include two categories: languages for smart contracts and general-purpose languages.
The Rust programming language is featured in both categories.
However, the use case of Rust as a smart contract language requires slightly more work for integrating the resulting code with the target blockchain infrastructure (see~\cref{sec:rust-extraction}).

Our trusted computing base (TCB) includes Coq itself, the quote functionality of MetaCoq and the pretty-printing to target languages.
While the erasure procedure for types is not verified, it does not affect the soundness of the pipeline (see discussion in \cref{sec:extraction}).

When extracting large programs, the performance of the pipeline inside Coq might become an issue.
In such cases, it is possible to obtain an OCaml implementation of our pipeline using the standard Coq extraction.
However, this extends the TCB with the OCaml implementation of extraction and the pre-processing pass, since the proof terms will not be generated and checked in the extracted OCaml code.

Our development is open-source and available in the GitHub repository~\url{https://github.com/AU-COBRA/ConCert/tree/journal-2021}.

\section{The MetaCoq Project}\label{sec:MetaCoq}
Since MetaCoq is integral to our work, we briefly introduce the project structure and explain how different parts of it are relevant to our development.
The MetaCoq project~\citep{Anand:TemplateCoq} consists of several subprojects aiming for formalising the meta-theory of Coq in Coq itself.
Apart from the meta-theory formalisation, MetaCoq provides meta-programming facilities allowing for manipulating the Coq code at the meta-level.
Below, we outline several parts of the projects that are the most relevant for the present work.

\paragraph*{Template Coq.}
This subproject adds meta-programming facilities to Coq.
That is, Coq definitions can be \emph{quoted} giving an AST of the original term represented as an inductive data type \icode{term} internally in Coq.
The \icode{term} type and related data types in Template Coq are very close to the actual implementation of the Coq kernel written in OCaml, which makes the quote/unquote procedures straightforward to implement.
This representation is suitable for defining various term-level transformations as Coq functions with the type \icode{term -> term}.
Eventually, the transformed AST can be \emph{unquoted} back to an ordinary Coq definition (provided that the resulting term is well-typed).
This functionality opens possibilities for many applications common to metaprogramming, such as syntactic translations, automatic instance derivation, and, in general, plugin development.
A \emph{plugin} is a program, written in OCaml that manipulates the syntactic representation of Coq terms directly.
Template Coq allows for writing such plugins in Coq itself, making it possible to verify the plugin's implementation.
The implementation of the quote/unquote functionality is written in OCaml, and Template Coq itself is a plugin.
From that point of view, Template Coq is a plugin for writing plugins in Coq.

The Template Coq metaprogramming facilities are used as the first step in our pipeline.
Given a (potentially verified and dependently typed) program in Coq, we can use \emph{quote} to obtain the program's AST that is then transformed, extracted, optimised and finally pretty-printed to one of the target languages (see~\cref{fig:pipeline}).

Let us give an example of the Template Coq representation of term.
We run the following command in Coq to obtain a quoted representation of a term \icode{fun x y : nat => x + y}:
\begin{lstlisting}
MetaCoq Quote Definition plus_nat_syn : term := (fun x y : nat => x + y).
\end{lstlisting}
\noindent%
The command adds the top-level declaration \icode{plus_nat_syn} to the current scope.
We can inspect the result by printing the definition that gives us the following result.
\begin{lstlisting}
tLambda {| binder_name := nNamed "x"; binder_relevance := Relevant |} (* binder info *)

  (* binder type *)
  (tInd {| inductive_mind := (MPfile ["Datatypes"; "Init"; "Coq"], "nat");
           inductive_ind := 0 |} [])

  (tLambda {| binder_name := nNamed "y"; binder_relevance := Relevant |} (* binder info *)

   (* binder type *)
    (tInd {| inductive_mind := (MPfile ["Datatypes"; "Init"; "Coq"], "nat");
             inductive_ind := 0|}[])

    (* body *)
    (tApp (tConst (MPfile ["Nat"; "Init"; "Coq"], "add") []) [tRel 1; tRel 0]))
\end{lstlisting}
\noindent%
From the code snippet, one can see that lambda-abstractions carry the domain type.
The subterm \icode{(MPfile ["Datatypes"; "Init"; "Coq"], "nat")} of type \icode{kername} represents a fully qualified name of the type, in this case it is \icode{Coq.Init.Datatypes.nat}.
The body of the function is an application.
Similarly to the OCaml implementation of the kernel, the application is n-ary, therefore the \icode{tApp} constructor accepts a term (the head of the application) and a list of arguments.
Template Coq uses the nameless representation of variables (de Bruijn indices), expressed as the \icode{tRel : nat -> term} constructor.

Apart from the vernacular commands (e.g.\ \icode{MetaCoq Quote Definition ...}, Template Coq features the \emph{template monad}, which is similar in spirit to the IO monad and allows for interacting with the Coq environment (quote, unquote, query and add new definitions, etc.).
We use the template monad in our pipeline for various whenever such interaction is required.
For example, we use it for implementing proof generating transformations (see~\cref{sec:proof-generating}).

\paragraph*{PCUIC.}
Predicative calculus of cumulative inductive constructions (PCUIC) is a variant of the calculus of inductive constructions (CIC) that serves as the underlying theoretical foundation of Coq.
In essence, PCUIC representations is a ``cleaned-up'' version of the kernel representation used in Template Coq.
In particular, compared to the Template Coq AST:
\begin{itemize}
\item PCUIC lacks type casts used to enforce a particular conversion mechanism in Coq.
\item PCUIC has the standard binary application, while it is n-ary in Template Coq (and in the OCaml implementation of the kernel).
\end{itemize}
MetaCoq features translation between the two representations.

The main purpose of PCUIC representation is to develop a formalisation of the meta-theory of Coq in Coq itself.
Various meta-theoretic results about PCUIC has been formalised in MetaCoq, including the verification of the type checker~\citep{CertErasure}.
We use the results related to reduction and typing in our development extensively.

\paragraph*{Verified erasure.}

One important part of the MetaCoq project that we build on is the verified erasure procedure.
The erasure procedure takes a PCUIC term as input and produces a term in \CICbox{}.
The meta-theory of PCUIC developed as part of MetaCoq is used extensively in erasure implementation and formalisation of the correctness results.
The erasure procedure is quite subtle and its formalisation is a substantial step towards the fully verified extraction pipeline.
We discuss the role and the details of MetaCoq's verified erasure in~\cref{sec:extraction}.

\section{The ConCert Framework}\label{sec:ConCert}
The present work builds on and extends the ConCert smart contract certification framework presented by the three authors of the present work at the conference Certified Programs and Proofs in January 2020~\citep{ConCert}.
In this section, we describe the overall structure of ConCert focusing on the parts, relevant for the present work, and extensions developed in~\citep{ConCert-extraction-testing} and in the present work.

The ConCert framework consists of the following layers.

\paragraph*{Embedding Layer.}
This layer features an embedding of smart contracts into Coq along with the proof of soundness of the embedding using the MetaCoq project~\citep{MetaCoq}.
Specifically, we show that the translation from the input language $\lambda_\mathrm{smart}$ to the PCUIC AST is computationally sound wrt.\ the weak call-by-value evaluation semantics.
The embedded contracts are available in the deep embedding (as ASTs) and in the shallow embedding (as Coq functions).
Having smart contracts as Coq functions facilitates the reasoning about their functional correctness properties.
The development features an example embedding of a functional smart contract language and several examples written directly in the deep embedding.

\paragraph*{Execution Layer.}
The execution layer provides a model that allows for reasoning on contract execution traces which makes it possible to state and prove temporal properties of interacting smart contracts.
In the functional smart contract model, the contracts consist of two functions:
\begin{itemize}
\item
  \begin{lstlisting}[basicstyle=\small]
init : Chain -> ContractCallContext -> Setup -> option State}
  \end{lstlisting}
  \vspace{-1em}
The initialisation function is called after the contract is deployed on the blockchain.
The first parameter of type \icode{Chain} represents the blockchain from a contract's point of view.
For example, a contract can access the current chain height, the current slot number (which can be used as timestamps) and so on.
Data about the current call to the contract (who calls the contact, the amount sent to the contract, etc.) is available through the second parameter of type \icode{ContractCallContext}.
\icode{Setup} is a user defined type that supplies custom parameters to the initialisation function.
The function might fail by returning \icode{None}. If the call succeeds, the function returns the initial value for the contract state.
\item
  \begin{lstlisting}[basicstyle=\small]
receive : Chain -> ContractCallContext -> State -> option Msg
          -> option (State * list ActionBody)
  \end{lstlisting}
  \vspace{-1em}
  This function represents the main functionality of the contract that is executed for each call to the contract.
  \icode{Chain} and \icode{ContractCallContext} are the same as for the \icode{init} function.
  The parameter of type \icode{State} represents the current state of the contract and \icode{Msg} is a user-defined type of messages that contract accepts.
  The result of the successful executions is a new state and a list of \emph{actions} represented with \icode{ActionBody}.
  The actions can be transfers, calls to other contracts (including the contract itself) and contract deployment actions.
\end{itemize}
\noindent%
Both \icode{receive} and \icode{init} are ordinary Coq functions, making it convenient to reason about.
However, as one can see from the signature of \icode{receive}, reasoning about the contract functions in isolation is not quite sufficient.
One contract call potentially emits more calls to other contracts, or to itself.
In an actual blockchain implementation, these calls would be handled by a \emph{scheduler}.
Our execution layer features a relational specification of a scheduler without committing to a particular order of processing messages in the list produced by each contract call.
The execution trace is defined as the following.
\begin{lstlisting}
ChainedList (Point : Type) (Link : Point -> Point -> Type) : Point -> Point -> Type :=
    clnil : forall p : Point, ChainedList Point Link p p
  | snoc : forall from mid to : Point,
           ChainedList Point Link from mid ->
           Link mid to -> ChainedList Point Link from to

 Definition ChainTrace := ChainedList ChainState ChainStep.
\end{lstlisting}
\noindent%
The definition of \icode{ChainTrace} is essentially a reflexive transitive closure of the proof-relevant relation \icode{ChainStep : ChainState -> ChainState -> Type}.
The steps in \icode{ChainStep} are \icode{step_block}, \icode{step_action} and \icode{step_permute}, which correspond to adding a block, executing an action (a transfer, a contract call, or a deployment of a new contract), and changing the order of action, scheduled for execution.
In Coq, these steps are represented as constructors of inductive family \icode{ChainStep}.

However, we also provide executable implementations of the specification that execute the outgoing call in depth-first or breadth-first order (see~\citep{Interactions} for more details).
The executable implementations are especially useful for techniques like property-based testing that we have explored in our previous work~\citep{ConCert-extraction-testing}.
\lstset{breaklines=false,keepspaces=false}

\paragraph*{Extraction Layer.}
The previous work on ConCert~\citep{ConCert} mainly concerns with the following use-case: take a smart contract in a functional smart contract language, \emph{embed} it into Coq and verify its properties.
This work shows how it is possible to verify a contract as a Coq function and then \emph{extract} it into a program in a functional smart contract language.
This layer represents an interface between the general extraction machinery we have developed and the use case of smart contracts.
Smart contract languages require a special approach in comparison with conventional extraction targets.
It is necessary to provide functionality for the integration of extracted smart contracts with the target blockchain infrastructure.
In practice, it means that we should be able to map the abstractions of the execution layer (contract's view of the blockchain, call context data) on corresponding components in the target blockchain.

Currently, all extraction functionality we have developed (regardless of the relation to smart contracts) is implemented in the extraction layer of ConCert.
In the future, we plan to separate the general extraction component from the blockchain-specific functionality.

\section{Extraction}\label{sec:general-extraction}
The Coq proof assistant comes with a dependently typed programming language Gallina that allows due to the language's rich type system to write programs together with their specifications in the style of \emph{certified programming} (see e.g.\ \citep{Chlipala:cpdt}).
Coq features a special universe of types for writing program specifications, the universe of propositions \icode{Prop}.
For example, the type \icode{\{n : nat | 0 < n \}} belongs to so-called \emph{subset types}, which are essentially a special case of a dependent pair type ($\Sigma$-type).
In this example, \icode{0 < n} is a proposition, i.e.\ it belongs to the universe \icode{Prop}.
Subset types allow for encoding many useful invariants when writing programs in Gallina.
An inhabitant of \icode{\{n : nat | 0 < n \}} is a pair with the first component being a natural number and the second component --- a \emph{proof} that the number is strictly greater than zero.
In the theory of Coq, subset types are represented as an inductive type with one constructor:
\begin{lstlisting}
  Inductive sig (A : Type) (P : A -> Prop) : Type :=
    exist : forall x : A, P x -> {x : A | P x}
\end{lstlisting}
\noindent%
where \icode{\{x : A | P x\}} is a notation for \icode{sig A P}.

The invariant represented by a second component can be used to ensure, for example, that division by zero never happens since we require that arguments can only be strictly positive numbers.
The proofs of specifications are only used to build other proofs and do not affect the computational behaviour of the program (apart from some exceptions called the singleton elimination principle).
The \icode{Prop} universe marks such computationally irrelevant bits.
Moreover, types appearing in terms are also computationally irrelevant.
For example, in System F this is justified by parametric polymorphism.
This idea is used in the Coq proof assistant to \emph{extract} the executable content of Gallina terms into OCaml, Haskell and Scheme.
The extraction functionality thus enables proving properties of functional programs in Coq and then automatically producing code in one of the supported languages.
The extracted code can be integrated with existing developments or used as a stand-alone program.

The first extraction using \icode{Prop} as a marker for computationally irrelevant parts of programs was introduced by~\cite{Paulin-Mohring:Extraction} in the context of the calculus of construction (CoC), which earlier versions of Coq were based on.
This first extraction targeted System F$\omega$, which can be seen as a subset of CoC, allowing one to get the extracted term \emph{internally} in CoC.
The current Coq extraction mechanism is based on the theoretical framework from a PhD thesis by~\cite{letouzey04}.
Letouzey extended the previous work of~\cite{Paulin-Mohring:Extraction} and adapted it to the full calculus of inductive constructions.
The target language of the current extraction is untyped, allowing to accommodate more features from the expressive type system of Coq.
The untyped representation has a drawback, however: the typing information is still required when extracting to statically typed programming languages.
To this end, Letouzey considers practical issues for implementing an efficient extraction procedure, including recovering the types in typed target languages and various optimisations.
The crucial part of the extraction process is the \emph{erasure} procedure that utilises the typing information to prune irrelevant parts.
That is, types and propositions in terms are replaced with $\square$~(a box).
Formally, it is expressed as a translation from \CIC{} (Calculus of Inductive Constructions) to \CICbox{} (an untyped version of \CIC{} with an additional constant $\square$).
The translation is quite subtle and is discussed in detail in~\citep{letouzey04}.
Letouzey also provides two (pen-and-paper) proofs that the translation is computationally sound: one proof is syntactic and uses the operational semantics and the other proof is based on the realisability semantics.
Computational soundness means that the original programs and the erased programs compute the same (in a suitable sense) value.

Having this in mind, we have identified two essential points:
\begin{itemize}
  \item The target languages supported by the standard Coq extraction do not include many new target languages, that represent important use cases (smart contracts, web programming).
  \item Since the extraction implementation becomes part of a TCB, one would like to mechanically verify the extraction procedure in Coq itself and the current Coq extraction is not verified.
\end{itemize}
\noindent%
Therefore, it is important to build a verified extraction pipeline in Coq itself that also allows for defining pretty-printers for new target languages.

Until recently, the proof of correctness of one of the essential ingredients, the erasure procedure, was only done on paper.
However, the MetaCoq project made an important step towards verified extraction by formalising the computational soundness of erasure~\citep[Section 4]{CertErasure}.
The MetaCoq's verified erasure is defined for predicative calculus of cumulative inductive constructions (PCUIC) a variant of \CIC{} that closely corresponds to the meta-theory of Coq (see~\cref{sec:MetaCoq} for a brief description of the project's structure and Section 2 of~\citep{CertErasure} for the detailed exposition of the calculus).
The result of the erasure is a \CICbox{} term, that is, a term in an untyped calculus.
On the other hand, integration with typed functional languages requires recovering the types from the untyped output of the erasure procedure.
In~\citep{letouzey04} this problem is solved by designing an erasure procedure for types and then using a modified type inference algorithm (based on the algorithm $\mathcal{M}$~\citep{AlgM:LeeYi}) to recover types and check them against the type produced by extraction.
Because the type system of Coq is more powerful than type systems of the target languages (e.g.\ Haskell or OCaml), not all the terms produced by extraction will be typable.
In this case, the modified type inference algorithm inserts type coercions forcing the term to be well-typed.
If we start with a Coq term the type of which is outside the OCaml type system (even without using dependent types), the extraction might have to resort to \icode{Obj.magic} in order to make the definition well-typed.
For example, the code snippet below
\begin{lstlisting}[language=Coq]
Definition rank2 : forall (A : Type), A -> (forall A : Type, A -> A) -> A
  := fun A a f => f _ a.
Extraction rank2.
\end{lstlisting}
gives the following output on extraction to OCaml:
\begin{lstlisting}
(** val rank2 : 'a1 -> (__ -> __ -> __) -> 'a1 **)
let rank2 a f = Obj.magic f __ a
\end{lstlisting}
\noindent%
These coercions are ``safe'' in the sense that they do not change the computational properties of the term, they merely allow to pass the type checking.

\lstset{keepspaces = true}
\subsection{Our Extraction}\label{sec:extraction}
The standard Coq extraction targets conventional general-purpose functional programming languages.
Recently, there has been a significant increase in the number of languages that are inspired by these, but due to the narrower application focus are different in various subtle details.
We have considered the area of smart contract languages (Liquidity and CameLIGO), web programming (Elm) and general-purpose languages with a functional subset (Rust).
They often pose more challenges than the conventional targets for extraction.
We have identified the following issues.
\begin{enumerate}
\item Most of the smart contract languages\footnote{At least, Simplicity, Liquidity, CameLIGO (and other LIGO languages), Love~\url{https://dune.network/docs/dune-node-next/love-doc/reference/love.html} (accessed 2021-08-05), Scilla and Sophia~\url{https://aeternity-sophia.readthedocs.io/} (accessed 2021-08-05).} and Elm do not offer a possibility to insert type coercions forcing the type checking to succeed.
\item The operational semantics of \CICbox{} has the following rule~\citep[Section 4.1]{CertErasure}: if $\wcbveval{\Sigma}{t_1}{\square}$ and $\wcbveval{\Sigma}{t_2}{\ident{v}}$ then $\wcbveval{\Sigma}{\bigl(t_1~~t_2 \bigr)}{\square}$, where $\wcbveval{-}{-}{-}$ is a big-step evaluation relation for \CICbox{}, $t_1$ and $t_2$ are \CICbox{} terms, and $\ident{v}$ is a \CICbox{} value.
  This rule can be implemented in OCaml using the unsafe features, which are, again, not available in most of our target languages.
  In lazy languages, this situation never occurs~\citep[Section 2.6.3]{letouzey04}, but most of the languages we consider follow the eager evaluation strategy.
\item Data types and recursive functions are often restricted. E.g.\ Liquidity, CameLIGO (and other LIGO languages) do not allow for defining recursive data types (like lists and trees) and limits recursive definitions to tail recursion on a single argument.\footnote{Some languages do not have this restriction, e.g.\ Love.}
  Instead, these languages offer built-in lists and finite maps (dictionaries).
\item Rust has a fully-featured functional subset, but being a language for systems programming, does not have a built-in garbage collector.
\end{enumerate}
Regardless of our design choices, the soundness of the extraction (given that terms evaluate in the same way before and after extraction) will not be affected.
In the worst case, the extracted term will be rejected by the type checker of the target language.

Let us consider in detail what the restrictions outlined above mean for extraction.
The first restriction means that certain types will not be extractable.
Therefore, our goal is to identify a practical subset of extractable Coq types and give the user access to transformations helping to produce well-typed programs.
The second restriction is hard to overcome, but fortunately, this situation should not often occur on the fragment we want to work.
Moreover, as we noticed before, terms that might give an application of a box to some other term will be ill-typed and thus, rejected by the type checker of the target language.
The third restriction can be addressed by mapping Coq's data types (lists, finite maps) to the corresponding primitives in a target language.
The fourth restriction applies only to Rust and means that we have to provide a possibility to ``plug-in'' a memory management implementation.
Luckily, Rust libraries contain various implementations one can choose from.

At the moment, we consider the formalisation of typing in target languages out of scope for this project.
Even though the extraction of types is not verified, it does not compromise run-time safety: if extracted types are incorrect, the target language's type checker will reject the extracted program.
If we followed the work in~\citep{letouzey04}, which the current Coq extraction is based on, giving guarantees about typing would require formalising of target languages type systems, including a type inference algorithm (possibly algorithm $\mathcal{M}$~\citep{AlgM:LeeYi}).
The type systems of many languages we consider are not precisely specified and are largely in flux.
Moreover, for the target languages without unsafe coercions, some of the programs will be untypeable in any case.
Therefore, we provide a pre-processing pass on the Template Coq representation, which allows one to apply certifying transformations in a compositional way.

On the other hand, for more mature languages (e.g.\ Elm) one can imagine connecting our formalisation of extraction with a language formalisation, proving the correctness statement for both the run-time behaviour and the typeability of extracted terms.

We extend the work on verified erasure~\citep{CertErasure} and develop an approach that uses a minimal amount of unverified code that can affect the soundness of the verified erasure.
Our approach adds an erasure procedure for types, verified optimisations of the extracted code and pretty-printers for several target languages.
The main observation is that the intermediate representation $\IR$ corresponds to the core of a generic functional language.
Therefore, our pipeline can be used to target various functional languages with transformations and optimisations applied generically to the intermediate representation.

Before introducing our approach, let us give some examples of how the verified erasure works and motivate the optimisations we propose.
\begin{lstlisting}
Definition sum_nat (xs : list nat) : nat := fold_right plus 0 xs.
\end{lstlisting}
produces the following \CICbox{} code:
\begin{lstlisting}
fun xs => Coq.Lists.List.fold_right ∎ ∎ Coq.Init.Nat.add O xs
\end{lstlisting}
Where the \lstinline{∎} symbol corresponds to computationally irrelevant parts.
The first two arguments to the erased versions of \icode{fold_right} are boxes, since \icode{fold_right} in Coq has two implicit arguments. They become visible if we switch on printing of implicit arguments:
\begin{lstlisting}
Set Printing Implicit.
Print sum_nat.
(* sum_nat = fun xs : list nat => @fold_right nat nat Init.Nat.add 0 xs
     : list nat -> nat *)
\end{lstlisting}
In this situation we have at least two choices: remove the boxes by some optimisation procedure, or leave the boxes and extract \icode{fold_right} in such a way that the first two arguments belong to some dummy data type.%
\footnote{There are two rules in the semantics of \CICbox{} that do not quite fit into the evaluation model of the languages we consider: pattern-matching on a box argument and having a box applied to some argument.
  The pattern-matching on a box case is addressed in the last version of MetaCoq and we include this optimisation in our pipeline.
  The applied box case requires implementing $\square$ as an argument consuming function, which is impossible in several of our target languages due to the absence of unsafe features.
  Therefore, we choose to implement $\square$ as the \icode{unit} type, potentially resulting in ill-typed programs after extraction.
  However, we have not encountered such cases in the examples we have considered.}
The latter choice cannot be made for some smart contract languages due to restrictions on recursion (\icode{fold_right} is not tail-recursive), therefore, we have to remap \icode{fold_right} and other functions on lists to the corresponding primitive functions.
In the following example,
\begin{lstlisting}
Definition square (xs : list nat) : list nat := map (fun x => x * x) xs.
\end{lstlisting}
\noindent
the \icode{square} function erases to
\begin{lstlisting}
fun xs => Coq.Lists.List.map ∎ ∎ (fun x => Coq.Init.Nat.mul x x) xs
\end{lstlisting}
\noindent
\sloppy The corresponding language primitive would be a function with the following signature: \icode{TargetLang.map: ('a -> 'b) -> 'a list -> 'b list}.
Clearly, there are two extra boxes in the extracted code that prevent us from directly replacing \icode{Coq.Lists.List.map} with \icode{TargetLang.map}.
Instead, we would like to have the following:
\begin{lstlisting}
  fun xs => Coq.Lists.List.map (fun x => Coq.Init.Nat.mul x x) xs
\end{lstlisting}
\noindent
In this case, we can provide a translation table to the pretty-printing procedure mapping \icode{Coq.Lists.List.map} to \icode{TargetLang.map}.
Alternatively, if one does not want to remove boxes, it is possible to implement a more sophisticated remapping procedure. It could replace \icode{Coq.Lists.List.map ∎ ∎} with \icode{TargetLang.map}, but it would require finding all constants applied to the right number of arguments (or $\eta$-expand constants) and only then replace them.
Remapping inductive types in the same style would involve more complications: constructors of polymorphic inductives will have an extra argument of a dummy type. This would require more complicated pretty-printing of pattern-matching in addition to the similar problem with extra arguments on the application sites.
\lstset{keepspaces = false}

By implementing the optimisation procedure we achieve two goals: remove redundant computations and make the remapping easier.
Removing the redundant computations is beneficial for smart contract languages since it decreases the computation cost in terms of \emph{gas}.
Users typically pay for calling smart contracts and the price is determined by the gas consumption.
That is, gas serves as a measure of computational resources required for executing a contract.
It is important to separate these two aspects of extraction: erasure (given by the translation \CIC{}\footnote{Note that by \CIC{} terms we mean in this section a particual version of it formalised in MetaCoq --- predicative calculus of cumulative inductive constructions (PCUIC)} $-->$ \CICbox{}) and optimisation of \CICbox{} terms to remove unnecessary arguments.
The optimisations we propose remove some redundant reductions, make the output more readable and facilitate the remapping to the target language's primitives.

Our implementation strategy of extraction is the following:
\begin{enumerate*}[label=(\roman*)]
\item take a term and erase it and its dependencies recursively to get an environment;
\item analyse the environment to find optimisable types and terms;
\item optimise the environment in a consistent way (e.g.\ in our $\IR$, the types must be adjusted accordingly);
\item pretty-print the result in the target language syntax according to the translation table containing remapped constants and inductives.
\end{enumerate*}

\subsubsection{Erasure for Types}\label{sec:erasure-for-types}
Let us discuss our first extension to the verified erasure presented in~\citep{CertErasure}, namely an \emph{erasure procedure for types}.
It is a crucial part for extracting to a \emph{typed} target language.
\begin{figure*}
  \footnotesize{%
    \begin{equation*}
      \begin{aligned}[t]
        \erasetypename{} & : \type{Ctx} -> \type{ECtx} -> \cic{\type{term}} -> \type{option~\mathbb{N}} \\ &
        -> \type{list~name} \times \boxty{\type{box\_type}}\\
          \erasetype{&\Gamma}{\Gamma_e}{\ident{\cic{t}}}{v_n} := \<let>~\ident{\cic{t'}} := \redbetaiotazeta~\Gamma~\ident{\cic{t}}~\<in>\\
        & \<let>~\ident{flag} := \flagoftype~\Gamma~\ident{\cic{t'}} \<in>\\
        & \<if>~(\<islogical>~\ident{flag})~\<then>~([\,],\TBox) ~ \<else>\\
        & \<match>~\ident{\cic{t'}}~\<with>\\
        & |~\cic{\overline{i}} => \<Ok> ([\,],\erasevar{\Gamma_e}{i})\\
        & |~\cic{\<tSort>} => ([\,],\TBox)\\
        & |~\cic{\<forall>~ a : A, B} =>\\
        & \quad \<let>~\ident{flag} := \flagoftype~\Gamma~\cic{A}~\<in>\\
        & \quad \<if>~(\<islogical>~\ident{flag})~\<then>\\
        & \qquad \<let>~(\ident{vs_\tau},\boxty{\tau}) := \erasetype{(\ident{\cic{A}}::\Gamma)}{(\<RelOther>::\Gamma_e)}{\cic{B}}{v_n}~\<in>\\
        & \qquad (\ident{vs_\tau}, \boxty{\square --> \tau})\\
        & \quad \<else>~\<if>~\neg(\<convtoarity>~\ident{flag})~\<then>\\
        & \qquad \<let>~(\ident{vs_\sigma}, \boxty{\ident{\sigma}}) := \erasetype{\Gamma}{\Gamma_e}{\ident{A}}{v_n}~\<in>\\
        & \qquad\<let>~(\ident{vs_\tau},\boxty{\tau}) := \erasetype{(\ident{\cic{A}}::\Gamma)}{(\<RelOther>::\Gamma_e)}{\cic{B}}{v_n}~\<in> \\
        & \qquad (\ident{vs_\tau}, \boxty{\sigma --> \tau})\\
        & \quad \<else>~\<let>~\ident{var} :=\\
          &\quad\qquad\<match>~v_n~\<with>\\
          &\qquad\qquad|~\<Some>~i => \<RelTypeVar>~i \quad |~\<None> => \<RelOther>\\
          &\quad\qquad\<end>~\<in>\\
          &\qquad\<let>~(\ident{vs_\tau},\boxty{\tau}) := \\
        & \qquad\quad\erasetype{(\ident{\cic{A}}::\Gamma)}{(\ident{var} :: \Gamma_e)}{\cic{B}}{(\incvar{}~v_n)}~\<in> \\
        &\qquad \<let>~\ident{vs} :=\\
        & \qquad\quad \<if>~(\isnone{}~v_n)~\<then>~\ident{vs_\tau}~\<else>~\cic{a}::\ident{vs_\tau}~\<in>\\
        &\qquad (\ident{vs}, \boxty{\square --> \tau})\\
        & \quad |~ \cic{\bigl(\ident{u}~~\ident{v}\bigr)} => \<let>~(\cic{\ident{hd}},\cic{\ident{args}}):= \decomposeapp~\cic{\left(\ident{u}~~\ident{v}\right)}~\<in>\\
        & \qquad \<let>~\boxty{\sigma} := \erasetypehead{\Gamma_e}{\cic{\ident{hd}}}~\<in>\\
        & \qquad\<if>~(\canhaveargs{}~\boxty{\sigma})~\<then>~([\,], \erasetypeapp{\Gamma_e}{\cic{\ident{args}}}{\ident{vs}}{\boxty{\sigma}})\\
        & \qquad\<else>~([\,], \boxty{\sigma})\\
        & \quad|~ \cic{\<tConst>} => ([\,],\boxty{\<tConst>})
          \quad|~ \cic{\<tInd>} => ([\,],\boxty{\<tInd>})
          \quad|~\_ => \boxty{\TAny}\\
        & \<end>
      \end{aligned}
       \begin{aligned}[t]\hspace{-1em}
         \erasetypeappname{} : ~ & \type{ECtx} -> \type{list~\cic{term}}\\
         & -> \boxty{\type{box\_type}}-> \boxty{\type{box\_type}}\\
        \erasetypeapp{&\Gamma_e}{\ident{\cic{args}}}{\boxty{\sigma}} :=\\
        & \<match>~\ident{\cic{args}}~\<with>\\
        & ~|~ [\,] => \boxty{\sigma}\\
        & ~|~ \cic{\ident{a}} ::\cic{\ident{args'}} =>\\
        & \qquad \<let>~\cic{\ident{A}} := \infer~\cic{\ident{a}}~\<in>\\
        & \qquad \<let>~\ident{flag} := \flagoftype~\Gamma~\ident{\cic{A}}~\<in>\\
        & \qquad \<let>~\boxty{\tau} :=\\
        & \qquad\quad \<if>~(\<islogical>~\ident{flag})~\<then>~\TBox\\
        & \qquad\quad  \<else>~\<if>~(\<issort> ~ \ident{flag})~\<then>\\
        & \qquad\qquad  \snd~(\erasetype{\Gamma}{\Gamma_e}{\ident{\cic{a}}}{\<None>})\\
        & \qquad\quad \<else>~\TAny~\<in>\\
        & \qquad \erasetypeapp{\Gamma_e}{\cic{\ident{args'}}}{\ident{vs}}
             {\boxty{\bigl({\sigma ~~ \tau}\bigr)}}\\
        & ~\<end>\\[.3em]
        \erasetypeheadname{} & : \type{ECtx} -> \type{\cic{term}} -> \boxty{\type{box\_type}}\\
          \erasetypehead&{\Gamma_e}{\ident{\cic{hd}}} :=\\
          & \<match>~\ident{\cic{hd}}~\<with>\\
        & ~|~ \cic{\overline{i}} => \<match>~\Gamma_e(i)~\<with>\\
        & \qquad\quad |~\<RelInductive>~\ident{\cic{\<tInd>}} => \ident{\boxty{\<tInd>}}\\
        & \qquad\quad |~\<RelTypeVar>~i => \boxty{\overline{i}}\\
        & \qquad\quad |~\_ => \boxty{\TAny}\\
        & \qquad\quad \<end>\\
        & ~|~ \cic{C} => \boxty{C}
          \quad|~ \cic{\<tInd>} => \boxty{\<tInd>}
          \quad|~ \_ => \boxty{\TAny}\\
        & ~\<end>\\[.3em]
        \erasevarname & : \type{ECtx} -> \mathbb{N} -> \boxty{\type{box\_type}}\\
          \erasevarname&~\Gamma_e~i :=\\
          & \<match>~\Gamma_e(i)~\<with>\\
        & ~|~\<RelTypeVar>~i => \boxty{\overline{i}}\quad |~\<RelOther> => \TAny{}
            \quad|~\<RelInductive>~\ident{\cic{\<tInd>}} => \ident{\boxty{\<tInd>}}\\
        & ~\<end>\\
      \end{aligned}
  \end{equation*}}\vspace{-10pt}
  \setlength{\belowcaptionskip}{-8pt}
  \caption{Erasure from \CIC{} types to $\boxty{\type{box\_type}}$}\label{fig:type-erasure}
\end{figure*}
Currently, the verified erasure of MetaCoq provides only a term erasure procedure which will erase any type in a term to a box.
For example, a function using the dependant pair type ($\Sigma$-type) might have a signature involving \icode{sig nat (fun n => n > 10)}, i.e. representing numbers that are larger than $10$.
Applying MetaCoq's \emph{term} erasure will erase this in its entirety to a box, while we are interested in a procedure that instead erases only the type scheme in the second argument: we expect type erasure to produce $\boxty{\type{sig} ~~ \type{nat} ~~ \square}$, where the square now represents an irrelevant type.

While our target languages have type systems that are Hindley-Milner based (and therefore, for which type inference is complete), we still need an erasure procedure for types to be able to extract inductive types.
Moreover, our target languages support various extensions, and their compilers may not always succeed to infer types.
For example, Liquidity has overloading of some primitive operations, e.g.\ arithmetic operations for primitive numeric types.
Such overloading introduces ambiguities that cannot be resolved by the type checker without type annotations.
CameLIGO requires writing even more types explicitly.
Thus, the erasure procedure for types is also necessary to produce such type annotations.
The implementation of this procedure is inspired by~\citep{letouzey04}.

We have chosen a semi-formal presentation in order to guide the reader through the actual implementation while avoiding clutter from the technicalities of Coq.
We use concrete Coq syntax to represent the types of \CIC\@.
We do not provide syntax and semantics of \CIC{}, for more information we refer the reader to Section 2 of~\citep{CertErasure}.
The types of $\IR$ are represented by the grammar below.
{\small\begin{align*}
  \boxty{\sigma},~\boxty{\tau} : \boxty{\type{box\_type}} & ::= \boxty{\overline{i}} ~|~ \boxty{\<tInd>} ~|~ \boxty{\<tConst>} ~|~ \boxty{\sigma~~\tau} ~|~\boxty{\sigma --> \tau} ~|~ \TBox ~|~\TAny
\end{align*}}
Here $\boxty{\overline{i}}$ represents levels of type variables, $\boxty{\<tInd>}$ and $\boxty{\<tConst>}$ range over names of inductive types and constants respectively.
Essentially, $\boxty{\type{box\_type}}$ represents types of an OCaml-like functional language extended with constructors $\TBox$ (``logical'' types) and $\TAny$ (types that are not representable in the target language).
Additionally, we use colours to distinguish between the \textcolor{cicterms}{\CIC{} terms} and the target \textcolor{boxtyterms}{erased types}.

\begin{coqdefinition}{\coqref{extraction/theories/Erasure.v:erase\_type\_aux}{5eb12f34fc29035eb0909ffdda523c4b70eaa462/extraction/theories/Erasure.v\#L718}}[Erasure for types]
  The erasure procedure for types is given by functions $\erasetypename{}$, $\erasetypeappname{}$ and $\erasetypeheadname{}$ in \cref{fig:type-erasure}.
\end{coqdefinition}

The \erasetypename{} function takes four parameters.
The first is a context $\type{Ctx}$ represented as a list of assumptions.
The second is an erasure context $\type{ECtx}$ represented as a sized list (vector) that follows the structure of $\type{Ctx}$; it contains either a translated type variable $\<RelTypeVar>$, information about an inductive type $\<RelInductive>$, or a marker for items in $\type{Ctx}$ that do not fit into the previous categories $\<RelOther>$.
The last two parameters represent terms of \CIC{} corresponding to types and an index of the next type variable.
The next type variable index is wrapped in the $\type{option}$ type, and becomes $\<None>$ if no more type variables should be produced.

The erasure function $\erasetypename{}$ returns a tuple consisting of a list of type variables and a $\boxty{\type{box\_type}}$.
In some cases both $\TBox$ and $\TAny$ can be removed from the extracted code by optimisations, although $\TAny$ might require type coercions in the target language.
Note also that types do not have binders, since they represent prenex-polymorphic types.
The levels of type variables are numbers counted from the root to the usage site starting from zero.
For example a signature of \icode{map : ('a -> 'b) -> 'a list -> 'b list} can be written as follows ($[a; b]$ is a context of type varibles for the type).
\[ ([a; b], \boxty{(\overline{0} --> \overline{1}) --> \mathtt{list}~\overline{0} --> \mathtt{list}~\overline{1}}) \]
The functions $\erasetypename{}$ and $\erasetypeappname{}$ are defined by mutual recursion.
The $\decomposeapp{}$ function returns the head of an application and a (possibly empty) list of arguments.
We use $\erasetypeheadname{}$ to erase the head and $\erasetypeappname{}$ to process all the arguments.
The $\canhaveargs{}$ analyses the given type of and returns $\mathtt{true}$ if it is the name of an inductive or a constant, and $\mathtt{false}$ otherwise.
We also make use of the destructuring let notation for tuples $\<let>~(a,b) := \ldots$ and projections $\fst$ and $\snd$.
In our implementation, we extensively use dependently typed programming, so the actual type signature of the functions in \cref{fig:type-erasure} also contain proofs that terms are well-typed.
The termination argument is given by a well-founded relation, since the erasure starts with $\beta\iota\zeta$-reduction using the $\redbetaiotazeta$ function and then later recurses on subterms of this.
Here $\beta$ is reduction of applied $\lambda$-abstractions, $\iota$ is reduction of $\<match>$ on constructors, and $\zeta$ is reduction of the $\<let>$ construct.
The $\redbetaiotazeta$ function reduces until the head cannot be $\beta\iota\zeta$-reduced anymore and then stops; it does not recurse on subterms.
This reduction function is defined in MetaCoq also by using well-founded recursion.
Due to the well-founded recursion we write $\erasetypename{}$ as a single function in our formalization by inlining the definitions of $\erasetypeappname{}$ and $\erasetypeheadname{}$; this makes the well-foundedness argument easier.

One of the advantages of implementing the extraction pipeline in Coq directly is that we can use the verified meta-theory of Coq in our development.
For example, since we define the erasure procedure for types as a total function that accepts only well-typed terms, we should be able to show that all the reduction machinery we use does not break the well-typedness of terms.
For that purpose, we use two results: the reduction function is sound with respect to the relational specification, and the subject reduction lemma, that is, reduction preserves typing.
We extensively use the \texttt{Equations} Coq plugin~\citep{Equations} in our development to help managing the proof obligations related to well-typed terms and recursion.

An important device used to determine erasable types (the ones we turn into the special target types $\TBox$ and $\TAny$) is the function $\flagoftype{}:~\type{Ctx} -> \type{\cic{term}} -> \type{type\_flag}$, where the return type $\type{type\_flag}$ is defined as a record with two fields: $\<islogical>$ and $\<convtoarity>$. The $\<islogical>$ field carries a boolean, while  $\<convtoarity>$ carries a proof or a disproof of convertibility to an arity.
For the purposes of the presentation in the paper, we treat $\<convtoarity>$ as a boolean value, while in the implementation we use the proofs carried by $\<convtoarity>$ to solve proof obligations for the definition of $\erasetypename{}$.

A type is an \emph{arity} if it is a (possibly nullary) product into a sort: $\forall \vec{\cic{a}} : \vec{\type{\cic{A}}}, \type{\cic{s}}$ for $\cic{\type{s}} = \cic{\type{Type}} ~|~ \cic{\type{Prop}}$ and $\vec{\cic{a}} : \vec{\type{\cic{A}}}$ a vector of (possibly dependent) binders and types.
Inhabitants of arities are \textit{type schemes}.

The predicate $\<issort>$ tells us if a given type is a \emph{sort}, i.e. \ $\cic{\type{Prop}}$ or $\cic{\type{Type}}$.
Sorts are always arities.
Therefore, we use $\<issort>$ that turns a proof of converitibility to an arity into a proof of convertivility to a sort (or returns $\<None>$ if it is not the case).
Finally, a type is \emph{logical} when it is a proposition (i.e.\ inhabitants are proofs) or when it is an arity into $\type{\cic{Prop}}$: $\forall \vec{\cic{a}} : \vec{\type{\cic{A}}}, \type{\cic{Prop}}$ (i.e. inhabitants are propositional type schemes).
As concrete examples, $\cic{\type{Type}}$ is an arity and a sort, but not logical.
$\cic{\type{Type}} -> \cic{\type{Prop}}$ is logical, an arity, but not a sort.
$\cic{\<forall>~A : \type{Type},~A -> A}$ is neither of the three.

The difference with our previous erasure procedure for types given in~\citep{ConCert-extraction-testing} is twofold.
First, we make the procedure total.
That means that it does not fail in the cases when it hits a non-prenex type, instead, it tries to do its best or emits $\TAny$ if no further options are possible.
In particular, we have improved the handling of arities that makes it possible to extract programs defined in terms of elimination principles.
For example one can define \icode{map} in the following way: \icode{list_rect (fun x => list B) [] (fun x _ rec => f x :: rec) xs}.
Where \icode{list_rect} is the dependent elimination principle for lists.
\begin{lstlisting}
list_rect : forall (A : Type) (P : list A -> Type),
  P [] ->
  (forall (a : A) (l : list A), P l -> P (a :: l)) ->
  forall l : list A, P l
\end{lstlisting}
\noindent%
Clearly, the type of \icode{list_rect} is too expressive for the target languages we consider.
However, it is still possible to extract a well-typed term for the definition of \icode{map} above.
The extracted type of \icode{list_rect} looks as follows.
\[ ([a;p], \boxty{\overline{1} --> (\overline{0} --> \mathtt{list}~\overline{0} --> \overline{1} --> \overline{1}) --> \mathtt{list}~\overline{0} --> \overline{1}}) \]

Second, we have introduced an erasure procedure for type schemes.
The procedure allows us to handle type aliases, that is, Coq definitions that being applied to some arguments return a type.
Type aliases are used quite extensively in the standard library.
For example, the standard finite maps \icode{FMaps} contain definitions like \icode{Definition PositiveMap.t : Type -> Type := PositiveMap.tree}.
In $\eta$-expanded form it is a function that take a type and returns a type: \icode{fun T => PositiveMap.tree T}.
Without this extension, we would not be able to extract programs that use such definitions.
\begin{coqdefinition}{\coqref{extraction/theories/Erasure.v:erase\_type\_scheme}{5eb12f34fc29035eb0909ffdda523c4b70eaa462/extraction/theories/Erasure.v\#L983}}[Erasure for type schemes]
  The erasure procedure for type schemes is given by two functions $\erasetypeschemename{}$ and $\erasetypeschemeetaname{}$ in~\cref{fig:type-scheme-erasure}.
\end{coqdefinition}
The signatures of $\erasetypeschemename{}$ and $\erasetypeschemeetaname{}$ are similar to $\erasetypename{}$ but we also add a new context $\type{ACtx}$ representing the type of a type scheme, which we call arity.
So, for an arity $\forall (a : A) (b : B) \ldots (z : Z), \type{Type}$, we have $\Gamma_a = [(a,A); (b,B);\ldots;(z,Z)]$.
The $\erasetypeschemename{}$ function reduces the term and then, if it is a lambda-abstraction, looks at the result of $\flagoftype$ for the domain type.
If it is a sort (or, more generally, an arity) it adds a type variable.
If the reduced term is not a lambda abstraction, we know that it requires $\eta$-expansion, since its type is $\forall (a' : A'), t$.
Therefore, we call $\erasetypeschemeetaname{}$ with the arity context $(a',A') :: \Gamma_a$.
A simple example of a type scheme is the following:
\begin{lstlisting}
Definition Arrow (A B : Type) := A -> B.
\end{lstlisting}
\noindent%
It erases to a pair consisting of a list of type variables and a $\boxty{\type{box\_type}}$:
\[ ([a;b], \boxty{\overline{0} --> \overline{1}}) \]
Type schemes that use dependent types can also be erased.
For example, one can create an abbreviation for the type of sized lists.
\begin{lstlisting}
Definition vec (A : Type) (n : nat) := {xs : list A | length xs = n}.
\end{lstlisting}
\noindent%
which gives us the following type alias
\[ ([a;n], \boxty{\type{sig}~(\type{list}~\overline{0})~~\TBox}) \]
\noindent%
where \icode{sig} corresponds to the dependent pair type in Coq given by the notation \icode{\{xs : list A | length xs = n \} := sig (list A) (fun xs => length xs = n)}.
The erased type can be further optimised by removing the occurrences of $\TBox$ and irrelevant type variables.

The two changes described above bring our implementation closer to the standard extraction of Coq and allow for more programs to be extracted in comparison with our previous work.
Returning $\TAny$ instead of failing creates more opportunities for target languages that support unsafe type casts.

Having defined the erasure procedure for types, we implement an erasure procedure for inductive definitions.
Bringing it all together with the verified erasure of MetaCoq and the erasure for type schemes, we can define a procedure that erases lists of global declarations, which are called \emph{global environments}.
We enrich the representation of global environments of the MetaCoq's erasure with the typing information we obtained using $\erasetypename{}$.
Each entry in the global environment is represented by the following inductive type.
\begin{lstlisting}
Inductive global_decl :=
| ConstantDecl : constant_body -> global_decl
| InductiveDecl : mutual_inductive_body -> global_decl
| TypeAliasDecl : option (list type_var_info * box_type) -> global_decl.
\end{lstlisting}
\noindent%
where \icode{constant_body} adds the constant's erased type (the \icode{cst_type} field), which is absent in the corresponding definition of MetaCoq's $\CICbox$:
\begin{lstlisting}
  Record constant_body :=
  { cst_type : list name * box_type;  cst_body : option term; }.
\end{lstlisting}
Moreover, \icode{mutual_inductive_body} is enriched with typing information as well.
We explicitly treat type aliases by having a separate entry \icode{TypeAliasDecl}, which corresponds to type schemes.
We call the representation above $\IR$ and use it as an intermediate representation.

\begin{figure*}
  \footnotesize{%
  \begin{equation*}
      \begin{aligned}[t]
        \erasetypeschemename{} & : \type{Ctx} -> \type{ECtx} -> \type{ACtx} -> \cic{\type{term}} -> \type{\mathbb{N}} \\
        & -> \type{list~name} \times \boxty{\type{box\_type}}\\
        \erasetypescheme{&\Gamma}{\Gamma_e}{[\,]}{\ident{\cic{t}}}{v_n} = ([\,], \snd~(\erasetype{\Gamma}{\Gamma_e}{\cic{t}}{\<None>})) \\
        \erasetypescheme{&\Gamma}{\Gamma_e}{(\ident{na'},\ident{A'})}{\ident{\cic{t}}}{v_n} = \\
        & \<let>~\cic{t'} := \redbetaiotazeta{}~\Gamma~\cic{t}~\<in>\\
        & \quad\<match>~\cic{t'}~\<with>\\
        & \quad|~\cic{\lambda~(a : A).b} =>\\
        & \qquad \<let>~\ident{flag} := \flagoftype~\Gamma~\ident{\cic{A}}~\<in>\\
        &\qquad \<let>~\ident{v_n'} := \<if>~(\<convtoarity>~\ident{flag})~\<then>~v_n+1\\
        &\qquad\qquad~\<else>~v_n~\<in>\\
        &\qquad \<let>~\ident{kind} := \<if>~(\<convtoarity>~\ident{flag})~\<then>~(\<RelTypeVar>~v_n)\\
        &\qquad\qquad~\<else>~\<RelOther>~\<in>\\
        &\qquad \<let>~(\ident{vs}, \boxty{\tau}) :=\\
        &\qquad\quad\erasetypescheme{(\cic{A} :: \Gamma)}{(\ident{kind} :: \Gamma_e)}{b}{\Gamma_a}{u}{v_n'}~\<in>\\
        &\qquad(v_n' :: \ident{vs}, \boxty{\tau})\\
        &\quad|~ \_ => \erasetypeschemeeta{\Gamma}{\Gamma_e}{(\ident{na'},\ident{A'}) :: \Gamma_a}{u}{t}\\
        &\quad\<end>
      \end{aligned}
      \begin{aligned}[t]
        \erasetypeschemeetaname{} & : \type{Ctx} -> \type{ECtx} -> \type{ACtx} -> \cic{\type{term}} -> \type{\mathbb{N}} \\
       & -> \type{list~name} \times \boxty{\type{box\_type}}\\
      \erasetypeschemeeta{&\Gamma}{\Gamma_e}{[\,]}{\ident{\cic{t}}}{v_n} = ([\,], \snd~(\erasetype{\Gamma}{\Gamma_e}{t}{\<None>})) \\
        \erasetypeschemeeta{&\Gamma}{\Gamma_e}{(\ident{na'},\ident{A'})}{\ident{\cic{t}}}{v_n} = \\
        & \<let>~\ident{flag} := \flagoftype~\Gamma~\ident{\cic{A}}~\<in>\\
        &\<let>~\ident{v_n'} := \<if>~(\<convtoarity>~\ident{flag})~\<then>~v_n+1\\
        &\qquad\qquad~\<else>~v_n~\<in>\\
        &\<let>~\ident{kind} := \<if>~(\<convtoarity>~\ident{flag})~\<then>~(\<RelTypeVar>~v_n)\\
        &\qquad\qquad~\<else>~\<RelOther>~\<in>\\
        &\<let>~\ident{tapp} := \cic{\bigl(\ident{(\lift{t}})~~\ident{\overline{0}}\bigr)}~\<in>\\
        &\<let>~(\ident{vs}, \boxty{\tau}) := \\
        \quad&\quad\erasetypescheme{(\cic{A} :: \Gamma)}{(\ident{kind} :: \Gamma_e)}{\ident{tapp}}{\Gamma_a}{u}{v_n'}~\<in>\\
        &(v_n' :: \ident{vs}, \boxty{\tau})
      \end{aligned}
  \end{equation*}}
  \caption{Erasure for type schemes}\label{fig:type-scheme-erasure}
\end{figure*}

\subsubsection{Optimising extracted code}\label{sec:optimisations}
Our second extension of the verified erasure is \emph{deboxing} --- a simple optimisation procedure for removing some redundant constructs (boxes) left after the erasure step.
First, we observe that removing redundant boxes is a special case of more general optimisation: dead argument elimination.
Informally it boils down to the equivalence $(\lambda x.~t)~v \sim t$ when $x$ does not occur free in $t$.
Here $\sim$ means that both sides evaluate to the same value.
Then, deboxing becomes a special case: $(\lambda A~x.~t)~\square \sim \lambda x.~t$.
From erasure, we know that the variable $A$ does not occur free in \icode{t}.\footnote{In our implementation we do not rely on this property and instead more generally remove unused parameters.}
Having in mind this equivalence, we implement in Coq a function with the following signature:
\vspace*{-1pt}
{\small\begin{align*}
  \dearg~:&~ \type{ind\_masks} -> \type{cst\_masks} -> \type{term} -> \type{term}
\end{align*}}%
The first two parameters are lookup tables for inductive definitions and for constants defining which arguments of constructors and constants are unused.
The information about unused arguments is represented using \emph{masks} --- lists of boolean values with \icode{true} denoting the unused arguments.
The type $\type{term}$ represents \CICbox{} terms.
The $\dearg$ function traverses the term and adjusts all applications of constants and constructors using the masks.

We define the following function that processes the definitions of constants:
{\small\begin{align*}
  \deargcst ~:&~ \type{ind\_masks} -> \type{cst\_masks} -> \type{constant\_body} -> \type{constant\_body}
\end{align*}}%
This function deargs the body using $\dearg$ and additionally removes lambda abstractions in correspondence to the mask for the current constant.
Note that, since the masks apply only to constants in the program, we only remove dead arguments of top-level functions: abstractions representing closures are untouched.
Additionally, as dearging removes arguments from the top-level function, we must adjust the type signatures produced by the type erasure correspondingly.
For example, for the constant \icode{Definition foo (n m k : nat) := n} we get a mask \icode{mask = [ false; true; true]} and the optimised constant \icode{Definition foo (n : nat) := n}

To generate the masks we implement an analysis procedure that finds dead parameters of constants and dead constructor arguments.
For arguments of constants, we check syntactically if they do not appear in the body, while for constructor arguments we find all unused arguments in pattern matches and projections across the whole program.
This is implemented as a linear pass over each function body that marks all uses of arguments and constructor arguments in that function.
As we noted above the erased arguments will be unused and therefore this procedure gives us a safe way of removing many redundant boxes (cf.\ \citep[Section 4.3]{letouzey04}).

The syntactic check is quite imprecise; for example, it will not remove a parameter if its only use is to be passed to another function in which it is also unused.
To deal with this the analysis and dearging procedure can be iterated multiple times, but since our main use of the dearging is to remove arguments that are erased, this is not necessary.

For definitions of inductive types, we define the function
{\small\begin{align*}
  \deargmib ~ :~& \type{mib\_masks} -> \mathbf{N} -> \type{one\_inductive\_body} -> \type{one\_inductive\_body}
\end{align*}}%
which adjusts the definition of one inductive's body of a (possibly) mutual inductive definition.
With $\deargcst$ and $\deargmib$, we can now define a function that removes arguments according to given masks for all definitions in the global environment:
{\small\begin{align*}
        \deargenv~:&~ \type{ind\_masks} -> \type{cst\_masks} -> \type{global\_env} -> \type{global\_env}
\end{align*}}%
Dearging is then done by first analyzing the environment to obtain $\type{ind\_masks}$ and $\type{cst\_masks}$ and then applying the $\deargenv$ function.

We prove dearging correct under several assumptions on the masks and the program being erased.

First, we assume that all definitions in the program are closed, which is a reasonable assumption given by typing.
Secondly, we assume that the masks are \emph{valid}, meaning that all removed arguments of constants and constructors should be unused in the program.
By unused we mean that the argument does not syntactically appear except for in the binder.
The analysis phase outlined above is responsible for generating masks that satisfy this condition, although currently, we do not prove this and instead recheck that the condition holds for the masks that were output.
Finally, we assume that the program is $\eta$-expanded according to all the masks: all occurrences of constructors and constants should be applied to the arguments that are supposed to be removed.
We implement a \emph{certifying} procedure that performs $\eta$-expansion and generates proofs that the expanded terms are equal to the original ones (see~\cref{sec:proof-generating}).
The erasure procedure is a pruning transformation, meaning that it does not remove abstractions or arguments in applications, it just replaces some terms with $\boxty{\Box}$.
Therefore, $\eta$-expanded terms are preserved by erasure.
We, however, have not formalised this result and currently validate the terms after erasure to ensure that they are applied enough.

Our Coq formalisation features a proof of the following soundness theorem about the $\dearg$ function.
\begin{coqtheorem}{\coqref{extraction/theories/OptimizeCorrectness.v:dearg\_correct}{5eb12f34fc29035eb0909ffdda523c4b70eaa462/extraction/theories/OptimizeCorrectness.v\#L2809}}[Soundness of dearging]\label{thm:dearg-sound}
    Let $\boxty{\Sigma}$ be a closed erased environment and $\boxty{t}$ a closed \CICbox{}-term such that $\boxty{\Sigma}$ and $\boxty{t}$ are valid and expanded according to provided masks.\\
    Then
    \vspace{-0.25\baselineskip}\[ \wcbveval{\boxty{\Sigma}}{\boxty{t}}{\boxty{v}} \]
    implies
    \vspace{-0.25\baselineskip}\[ \wcbveval{\deargenv(\boxty{\Sigma})}{\dearg(\boxty{t})}{\dearg(\boxty{v})} \]
    where dearging is done using the provided masks.
\end{coqtheorem}%
\noindent
Here $\wcbveval{\boxty{-}}{\boxty{-}}{\boxty{-}}$ denotes the big-step call-by-value evaluation relation of \CICbox{} terms\footnote{The relation is part of MetaCoq. We contributed to fixing some issues with the specification of this relation.} and values are given as a subset of terms.
The theorem ensures that the dynamic behaviour is preserved by the optimisation function.
This result, combined with the fact that the erasure from \CIC{} to \CICbox{} preserves dynamic behaviour as well, gives us guarantees that the terms that evaluate in \CIC{} will be evaluated to related results in \CICbox{} after optimisations.

\cref{thm:dearg-sound} is a relatively low-level statement talking about the dearging optimisation that is used by our extraction.
The extraction pipeline itself is more complicated and works as outlined at the end of \cref{sec:extraction}: it is provided a list of definitions to extract in a well-typed environment and recursively erases these and their dependencies (see the full pipeline in \cref{fig:pipeline}).
Note that only dependencies that appear in the erased definitions are considered as dependencies; this typically gives an environment that is substantially smaller than the original.
Once the procedure has produced an environment, the environment is analysed to find out which arguments can be removed from constructors and constants, and finally, the dearging procedure is invoked.

MetaCoq's correctness proof of erasure requires the full environment to be erased.
Since we only erase dependencies we prove a strengthened version of the erasure correctness theorem that is applicable for our case.
Combining this with \cref{thm:dearg-sound} allows us to obtain a statement about the extraction pipeline (starting from the PCUIC representation and excluding the pretty-printing).

\begin{coqtheorem}{\coqref{extraction/theories/ExtractionCorrectness.v:extract\_correct}{5eb12f34fc29035eb0909ffdda523c4b70eaa462/extraction/theories/ExtractionCorrectness.v\#L56}}[Soundness of extraction]\label{thm:extract-sound}
    Let $\cic{\Sigma}$ be a well-typed axiom-free environment and let $\cic{C}$ be a constant in $\cic{\Sigma}$.
    Let $\boxty{\Sigma'}$ be the environment produced by successful extraction (including optimisations) of $\cic{C}$ from $\cic{\Sigma}$.
    Then, for any unerasable constructor \cic{Ctor}, if
    \vspace{-0.25\baselineskip}\[ \wcbvevalpcuic{\cic{\Sigma}}{\cic{C}}{\cic{Ctor}} \]
    it holds that
    \vspace{-0.25\baselineskip}\[\wcbveval{\boxty{\Sigma'}}{\boxty{\<tConst>}}{\boxty{Ctor}}\]
\end{coqtheorem}%
\noindent
Here $\wcbvevalpcuic{\cic{-}}{\cic{-}}{\cic{-}}$ denotes the big-step call-by-value evaluation relation for CIC terms.
Informally, the above statement can be specialised to say that any program computing a boolean value will compute the same value after extraction.
Of course, one still has to keep in mind that the pretty-printing step of the extracted environment is not verified and the discrepancies of \CICbox{} and the target language's semantics as we outlined in \cref{sec:extraction}.

While the statement does not say anything about constructor applications,\footnote{It is hard to give an easily understandable statement since dearging removes applications.} it does informally generalise to any value that can be encoded as a number, since it can be used to show that each bit of the output will be the same.

One of the premises of \cref{thm:extract-sound} is that the environment is axiom-free, which is required for the soundness of erasure as stated in MetaCoq and adapted in our work.
In general, we cannot say anything about the evaluation of terms once axioms are involved.
One possible way of fixing this issue is by following the semantic approach as in Section 2.4 of~\citep{letouzey04}.

While dearging subsumes deboxing we cannot guarantee that our optimisation removes all boxes even for constants applied to all logical arguments due to cumulativity.%
\footnote{By cumulativity we mean subtyping for universes, i.e. \icode{A : Type}$_i$ is also \icode{A : Type}$_{i + 1}$ for any $i$.
Therefore, if a function takes an argument \icode{A : Type}, we can pass \icode{Prop}, since it is at the lowest level of the universe hierarchy.}
E.g.\ for \icode{@inl Prop Prop True : sum Prop Prop} it is tempting to optimise the extracted version \icode{inl ∎ ∎ ∎} into just \icode{inl}, but the optimised definition of the \icode{sum} type will still have the \icode{inl} constructor that takes one argument, because its type is \icode{inl : forall A B : Type, A -> A + B} and the argument \icode{A} is, in general, relevant for computation.

As mentioned previously, the dearging of functions removes parameters which means that it must also adjust the type signatures of those functions.
In addition to this adjustment of type signatures, we also do a final pass to remove logical inductive type parameters.
This step is completely orthogonal to the dearging of terms and serves only to remove useless type parameters.
This does not affect the dynamic semantics, but mistakes in it might mean that the code does not type-check in the target language.

For a concrete example, sigma types are defined in Coq as
\begin{lstlisting}
Inductive sig (A : Type) (P : A -> Prop) :=
  exist : forall x : A, P x -> sig A P
\end{lstlisting}%
In the constructor, \icode{P} is a type scheme while the argument of type \icode{P x} is a proof, so these are erased by type erasure, resulting in the type $\boxty{\type{A} --> \square --> \type{sig} ~~ \type{A} ~~ \square}$.
The analysis will show that the proof argument is never used since any use is also erased.
This means the constructor is changed to $\boxty{\type{A} --> \type{sig} ~~ \type{A} ~~ \square}$ as part of the dearging process, and any use of this constructor in a function (e.g. for pattern matching, or to construct a value) is similarly adjusted.
Finally, removal of logical type parameters means that the type parameter $P$ is completely removed from the type, giving the final constructor type as $\boxty{\type{A} --> \type{sig} ~~ \type{A}}$.
Function signatures using \icode{sig} are also adjusted correspondingly, having the $P$ argument removed.

After applying the optimisations we pretty-print the optimised code to several functional languages.
We discuss issues related to extraction to Liquidity and CameLIGO in \cref{sec:liquidity-cameligo-extraction}, to Elm in \cref{sec:elm-extraction}, and to Rust in \cref{sec:rust-extraction}.

\begin{figure}
  \begin{center}
    \begin{minipage}{0.8\textwidth}
      \begin{lstlisting}
Definition storage := Z.
Inductive msg := Inc (_ : Z) | Dec (_ : Z).
Program Definition inc_counter (st : storage) (inc : {z : Z | 0 <? z}) :
  {new_st : storage | st <? new_st} := st $+$ inc. (* proof omitted *)
Program Definition dec_counter (st : storage) (dec : {z : Z | 0 <? z}) :
  {new_st : storage | new_st <? st} := st $-$ dec. (* proof omitted *)
Definition my_bool_dec := Eval compute in bool_dec.

Definition counter (msg : msg) (st : storage)
  : option (list operation * storage) :=
  match msg with
  | Inc i => match (my_bool_dec (0 <? i) true) with
    | left h => Some ([], proj1_sig (inc_counter st (exist _ i h)))
    | right _ => None
    end
  | Dec i => match (my_bool_dec (0 <? i) true) with
    | left h => Some ([], proj1_sig (dec_counter st (exist _ i h)))
    | right _ => None
    end
  end.
      \end{lstlisting}
    \end{minipage}
  \end{center}
\vspace{-15pt}
\setlength{\belowcaptionskip}{-8pt}
\caption{The \texttt{counter} contract}\label{lst:counter}
\end{figure}

\subsubsection{Handling absurd cases}\label{sec:false-elim}

Our approach should be able to handle the cases when Coq programs contain some unreachable code, originating from provably impossible cases.
As an example let us consider the following program in Coq.
\begin{lstlisting}
  Program Definition safe_head {A} (non_empty_list : {l : list A | length l > 0}) : A :=
    match non_empty_list as l' return l' = non_empty_list -> A  with
    | [] => fun _ => False_rect _ _
    | hd :: tl => fun _ => hd
    end eq_refl.
\end{lstlisting}
\noindent%
The type of the program ensures that one always can take the first element of the input list.
In the body of the program, we have to deal with two cases for the given list.
Clearly, we should never hit the empty list case.
Therefore, we use \icode{False_rect : forall P : Type, False -> P} that allows us to construct anything, provided we have a contradiction at hand.
Using the \icode{Program} tactic we construct such proof from the fact that \icode{length [] > 0} is in fact a contradiction.
In Coq, this definition gives us a total function \icode{safe_head}, which we then can use in other definitions, provided that we can construct an element of \icode{\{l : list A | length l > 0\}}.
For example, we can use it in the following program.
\begin{lstlisting}
  Program Definition head_of_repeat_plus_one {A} (n : nat) (a : A) : A
    := safe_head (repeat a (1+n)).
  Next Obligation. intros. cbn. lia. Qed.
\end{lstlisting}
\noindent%
However, in the extracted code, \icode{safe_head} must return some value of the appropriate type in the case of an empty list.
It can be done in different ways, depending on the features available in a target language.
One way of doing this would be to throw an exception in languages that support this kind of side effect.
E.g.\ the standard Coq extraction to OCaml uses \icode{assert false} for that purpose.
For the languages that do not support exceptions, we can use non-termination for the same purpose.
Since all the target languages we consider are eager, we should be a bit careful in how we represent such constructs.
Particularly, we need to guard the code that throws an exception or makes a non-terminating recursive call with a lambda abstraction.
Therefore, we can add a constant \icode{false_elim : unit -> a}, that works for any type \icode{a} and use this constant once we encounter pattern-matchings on any empty inductive type (an inductive type with no constructors, e.g.\ \icode{False}).

Another issue, related to extraction of \icode{False_rect} is how we apply our dearging optimisation.
By default, all the arguments of \icode{False_rect} will be removed by the optimisation.
Then, at the pretty-printing stage, the body of \icode{False_rect} will be replaced with a call \icode{false_elim ()}, which will be immediately evaluated in eager languages.
Therefore, following~\citep{letouzey04}, we adopt the following strategy.
At the analysis stage, if all the arguments of a constant are logical (i.e.\ of type $\boxty\TBox$ or $\boxty\TAny$) we generate a mask that keeps one argument, guarding the constant's body by a lambda abstraction.

We remark on how the pattern-matching on empty types is implemented in our targets in the corresponding sections.

\subsubsection{The Counter Contract}
As an example, let us consider a simple smart contract represetned as a Gallina function.
The state of the contract is an integer number and it accepts increment and decrement messages (\cref{lst:counter}, \coqref{extraction/examples/CounterSubsetTypes.v}{journal-2021/extraction/examples/CounterSubsetTypes.v}).
The main functionality is given by the two functions \icode{inc_counter} and \icode{dec_counter}. We use subset types to encode the functional specification of these functions.
E.g.\ for \icode{inc_counter} we encode in the type that the result of the increment is greater than the previous state given a positive increment.
Subset types are represented in Coq as dependent pairs ($\Sigma$-types).
For example a positive integer is encoded as \icode{\{z : Z | 0 <? z\}}, where the second component is a proposition \icode{0 <? i = true} (we use an implicit coercion from booleans to propositions).
Similarly, we encode the specification \icode{dec_counter}.
The \icode{counter} function validates the input and provides a proof that the input satisfies the precondition (of being positive).
The functions \icode{inc_counter} and \icode{dec_counter} are defined only for positive increments and decrements, therefore, we do not need to validate the input again.
Note that in order to construct an inhabitant of \icode{positive}, we use the decidability of equality for booleans \icode{bool_dec : forall b1 b2 : bool, \{b1 = b2\} + \{b1 <> b2\}} that gives us access to the proof of \icode{0 <? i}.
We will use the example from \cref{lst:counter} in subsequent sections for showing how it can be extracted to concrete target languages.

\lstset{basicstyle=\scriptsize}
\begin{figure*}
  \begin{subfigure}[t]{.47\textwidth}
    \begin{lstlisting}
type 'a sig_ = 'a
let exist_ a = a
type coq_msg = Coq_Inc of int | Coq_Dec of int
type storage = int
type coq_sumbool = Coq_left | Coq_right

let coq_inc_counter (st : storage) (inc : int sig_) =
     exist_ (addInt st ((fun x -> x) inc))
     ...

let coq_counter (msg : coq_msg) (st : storage) =
match msg with
  Coq_Inc i ->
  (match coq_my_bool_dec (ltInt 0 i) true with
    Coq_left  ->
    Some ([], ((fun x -> x)
       (coq_inc_counter st (exist_ (i)))))
  | Coq_right  -> None)
| Coq_Dec i -> ...
  | Coq_right  -> None)
    \end{lstlisting}
    \vspace{-5pt}
    \caption{Liquidity}\label{fig:extracted-liquidity}
  \end{subfigure}
  \begin{subfigure}[t]{.47\textwidth}
    \begin{lstlisting}
type 'a sig_ = a
let exist_ a = a
type coq_msg = Coq_Inc of int | Coq_Dec of int
type storage = int
type coq_sumbool = Coq_Left | Coq_Right
let coq_Transaction_none  = ([]: (operation) list)

let coq_inc_counter (st : storage) (inc : int sig_) =
  exist_ (addInt st ((fun (x: int sig_) -> x) inc))

...

let coq_counter (msg : coq_msg) (st : storage) =
match msg with
  Coq_Inc (i) ->
  (match coq_bool_dec true (ltInt 0 i) with
    Coq_Left -> (Some
    (coq_Transaction_none,
    ((fun (x: int sig_) -> x)
      (coq_inc_counter st (exist_ (i))))))
  | Coq_Right  -> (None: (operation list * storage) option))
| Coq_Dec (i) -> ...
    \end{lstlisting}
    \vspace{-5pt}
    \caption{CameLIGO}\label{fig:extracted-cameligo}
  \end{subfigure}
  \setlength{\abovecaptionskip}{5pt}
  \setlength{\belowcaptionskip}{-8pt}
  \caption{Extracted code.}\label{fig:extracted-code}
\end{figure*}

\lstset{basicstyle=\footnotesize}

\subsection{Proof-generating transformations}\label{sec:proof-generating}
The optimisation pass in our pipeline (see \cref{fig:pipeline}) expects constants and constructors to be applied to all logical arguments in order to be valid.
Moreover, some constants have types that are too expressive for the target languages that can make the extracted programs untypable.
However, the constants can be specialised in a way that the extracted code is well-typed.
In order to ensure that our input satisfies these and some other technical requirements, we apply transformation passes at the very beginning of our pipeline --- at the Template Coq level (see \cref{fig:pipeline}).
These transformation passes are implemented as unverified functions transforming the Template Coq AST.
In order to ensure that the passes are computationally sound, we apply the \emph{certifying} approach to transformation.
It is similar to how certifying compilers are used to produce proof-carrying code~\citep{Necula:ProofCarryingCode}.
The overall idea is that the transformation produces a new program (in our case it is the same language) and a proof term that the desired property is preserved by the transformation.
Each transformation in the Template Coq part of the pipeline has the following type
\begin{lstlisting}
  transform: global_env -> Result global_env string
\end{lstlisting}
\noindent%
Where \icode{global_env} is the Template Coq global environment (list of top level declarations), \icode{Result} is a error monad.
Given a list of transforms, we can compose them using the fact that $\type{Result}$ is a monad.
In fact, we can reuse the same way of composing transformation for different passes in our pipeline and define a common type of transformations as
\icode{Definition Transform (A : Type) := A -> result A string.}
As a result, we define the composition of transformation in the usual monadic way.

After successfully completing all the transformations, we can generate proofs that the definitions we transformed behave in the same way as the originals.
All the transformations we have considered have one property in common: they produce terms that are definitionally equal to the originals.
Definitional equality in Coq means that the two terms are \emph{convertible}, i.e.\ equivalent with respect to $\beta\delta\iota\zeta$-reduction, $\eta$-expansion and irrelevant terms\footnote{See more about the conversion mechanism in Coq's manual: \url{https://coq.inria.fr/refman/language/core/conversion.html}. Accessed: 2021-07-23}.
Where $\beta$ and $\eta$ are standard and $\delta$ means constant unfolding, $\iota$ --- reduction of \icode{match} on a constructor, $\zeta$ --- \icode{let .. in} reduction.
From the computational point of view, convertible terms represent the same program.
The fact that the terms are convertible gives us a simple way of generating the correctness proofs.
Let \icode{transform} be a transformation function and \icode{t0 : A} a term.
If \icode{transform t0 = Ok t1}, i.e.\ the application of this function to \icode{t0} succeeds with some transformed term \icode{t1}, we can construct the following proof term:
\begin{lstlisting}
  @eq_refl A t1 : t0 = t1
\end{lstlisting}
\noindent%
This proof term shows that we can prove that the two terms \icode{t0} and \icode{t1} are equal in the theory of Coq.
Moreover, this term is well-typed only if \icode{t0} and \icode{t1} are convertible.

We use the following approach to generating the proof terms:
\begin{itemize}
 \item Given a definition \icode{def}, quote it along with all the dependencies, producing a global environment $\Sigma_0$ with quoted terms.
 \item Apply the composed transformations to all elements in the original global environment $\Sigma_0$ and get the transformed environment $\Sigma_1$.
 \item For each constant from $\Sigma_0$ find a corresponding constant in $\Sigma_1$.
 \item If a constant is found, compare the constant bodies for \emph{syntactic} equality (it is possible since we operate in meta-theory).
   In case the bodies are not equal --- add (unquote) a new definition from $\Sigma_1$ to the current scope; if they are equal, or constant not found --- do nothing.
 \item If \icode{def} or its dependencies were affected by the transformation, generate a proof term and add (unquote) it to the current scope.
\end{itemize}

The certifying approach is quite flexible wrt.\ changes and additions of new passes since no modifications of proofs are required, provided that the passes preserve convertibility.
This is a big advantage in our setting when fine-tuning of the transformations is required for achieving the desired result (see, for example, the inlining transformation below).
Potentially, the pass can be extended with more general optimising transformations like partial evaluation.

Below, we describe transformations currently implemented in our framework.

\paragraph*{$\eta$-expansion \coqref{extraction/theories/CertifyingEta.v}{journal-2021/extraction/theories/CertifyingEta.v}.}
The idea is to find partially applied constants (or constructors) and expand them by introducing lambda-abstractions.
For example for a term \icode{let f := fun n => add n in f 0 0} would be expanded (if we demand full $\eta$-expansion) to \icode{let f := fun n m => add n m in f 0 0}.
The extent to which the expansion is performed is controlled by lookup tables mapping the names of constants (or constructor information) to a number, indicating the number of arguments that should be added, and the constant's (constructor's) type.
The typing information is required for introducing the lambda-abstractions since the Template Coq unquote functionality expects a fully specified term, and all binders typically have explicit types in the AST.
Calling the type checker would introduce too much overhead, therefore, we keep the required information in the lookup tables.
The transformation is mostly standard but requires a bit of care when dealing with types of lambda-abstractions.
Let us consider an example, writing all relevant types explicitly.
For the following code \icode{let f : list nat -> list nat := @cons nat 0 in f []}.
Our expansion table will contain the type of \icode{cons : forall \{A : Type\}, A -> list A -> list A}.
In order to introduce a lambda-abstraction, we need to know the type of the last argument of \icode{cons}.
Therefore, we need to specialise the type of \icode{cons} wrt.\ the arguments it is applied to.
We do so by substituting the arguments, to which the constant or a term is applied, into the term's type.

Since our main use case for $\eta$-expansion is to ensure that constants and constructors are applied to all logical arguments,
we use the masks generated by the analysis phase for optimisations (see \cref{sec:extraction}) to compute to which extent constants and constructors should be $\eta$-expanded.

Since $\eta$-equality is part of Coq's conversion mechanism, the resulting terms will be convertible to the originals.

\paragraph*{Expansion of \icode{match} branches.}
This transformation is tightly related to the representation of the \icode{match} construct in Coq.
The branches are represented as a list with each position corresponding to a constructor of the type of the discriminee.\footnote{By discriminee we mean a term on which the pattern-matching is performed.}
Each element of the list of branches is a pair with the first component being a number of a constructor's arguments, and the second --- a term, that can be applied to the number of arguments, specified in the first component.
The second component might be not $\eta$-expanded.
Let us consider a simple (contrived) example.
\begin{lstlisting}
 Definition match_list_id (xs : list nat) : list nat :=
 match xs with
 | [] => []
 | cons x xs => cons x xs
 end.
\end{lstlisting}
\noindent%
The internal representation of the branches is a list, admitting different ways of representing the second branch.
For example, it is perfectly fine to just use the \icode{cons} constructor applied only to the type of elements, but not to the other two arguments.
This list looks as follows in term of the AST constructors (we abbreviate the MetaCoq representation of the list of natural numbers as \icode{LIST} and the type of natural numbers as \icode{NAT}).
\begin{lstlisting}
  [(0, tApp (tConstruct LIST 0 []) [NAT]);
   (2, tApp (tConstruct LIST 1 []) [NAT])]
\end{lstlisting}
\noindent%
The pretty-printing procedure expects that all the branches start with lambdas if the corresponding patterns have arguments.
This invariant makes it possible to print the patterns in the usual way, with the top lambda-abstractions becoming pattern variables.
Therefore, we would like to expand the second branch, so it has the following shape (we abbreviate the binder information for lambda-abstractions as \icode{X} and \icode{XS}):
\begin{lstlisting}
tLambda X NAT
  (tLambda XS LIST
    (tApp (tConstruct LIST 1 []) [NAT; tRel 1; tRel 0])))])
\end{lstlisting}
\noindent%
Or, written in the concrete syntax \icode{fun (x : nat) (xs : list nat) => cons x xs}.

In most cases, writing a program in Coq does not lead to unexpanded representation of branches, but we have noticed that certain automatically generated definitions, like eliminators, might contain branches that are not expanded enough.
That can happen for automatically generated definitions.
One example of such a definition is \icode{sig_rect}, an eliminator for the \icode{sig} type from the standard library of Coq.
Without expansion, such definitions would prevent us from using our extraction pipeline.

The implementation of the branch expansion is similar to the $\eta$-expansion pass with one subtlety.
As we have noted before, we need to specify types for each binder introduced by lambda-abstractions.
Getting the information about the type of branches is quite complicated and with the current representation of branches in Template Coq would require running type inference.
Instead, we use a recent feature of Template Coq, called \emph{holes}.
Holes in Coq are represented by so-called existential variables, that can be manipulated by tactics and instantiated by the elaboration mechanism.
In our case, the surrounding context provides enough information for these variables to be instantiated.
Implementation-wise, due to similarities with the ``regular'' $\eta$-expansion, the passes are defined together.

\paragraph*{Inlining \coqref{extraction/theories/CertifyingInlining.v}{journal-2021/extraction/theories/CertifyingInlining.v}.}
The motivation for having an inlining pass is that some definitions that are not typable in the extracted code, become typable after inlining and specialising the bodies.
Inlining also helps to overcome some potential performance issues.
We have two common examples of this kind.
\begin{itemize}
\item Dependent eliminators.
  The code produced after extraction might be not typable because the original type is more expressive than prenex polymorphism in our target languages.
  Languages like CameLIGO do not support polymorphism at all.
  Moreover, using eliminators like \icode{bool_rect} (non-dependent version of it is essentially \icode{if_then_else}) is impractical, because the target languages use the call-by-value evaluation strategy.
  Therefore, evaluating expressions like \icode{bool_rect _ branch1 branch_2 cond} will effectively lead to evaluating both branches regardless of the condition, while we would like it to behave similarly to \icode{if_then_else}.
  After inlining, \icode{bool_rect} unfolds to pattern-matching and behaves as expected.
 \item General definition of a monad.
   The definition of a monad uses rank-2 polymorphism, which, again, goes beyond the supported types in the target languages.
   But inlining concrete instances of \icode{bind} and \icode{return} allows us to avoid this issue and continue using high-level abstraction in Coq while extracting the well-typed code.
\end{itemize}
In our framework, the inlining function has the following signature.
\begin{lstlisting}
  inline_in_env : (kername -> bool) -> global_env -> global_env
\end{lstlisting}
\noindent%
The first argument is a function indicating which constants should be inlined.
The second argument is the list of top-level declarations.
Apart from just inlining the bodies of specified constants, we also perform $\iota$ and $\beta$-reductions.
The extent to which the term is reduced is determined empirically from the applications to extraction.
Clearly, since inlining is $\delta$-reduction, accompanied with $\iota$- and $\beta$-reductions, the resulting terms are convertible to the original ones since all these reductions are part of Coq's conversion mechanism.

\subsection{Extracting to Liquidity and CameLIGO}\label{sec:liquidity-cameligo-extraction}
Liquidity is a functional smart contract language for the Tezos and Dune blockchains inspired by OCaml.
It compiles to Michelson\footnote{\url{https://tezos.gitlab.io/active/michelson.html}. Accessed 2021-07-21} --- a stack-based functional core language supported directly by the blockchain, developed by Tezos.

LIGO is another functional smart contract language for Tezos that compiles to Michelson.
LIGO has several concrete syntaxes: PascaLIGO, ReasonLIGO, and CameLIGO.
We target the CameLIGO syntax due to its similarity with Coq.

Compared to a conventional functional language, Liquidity and CameLIGO have many restrictions, mostly inherited from Michelson.
Hence, we present the key issues when extracting to these languages in a collapsed manner.

In both Liquidity and CameLIGO, data types are limited to non-recursive inductive types, and support for recursive definitions is limited to tail recursion on a single argument.%
\footnote{We reported the restrictions to the developers:
  recursive functions \url{https://github.com/OCamlPro/liquidity/issues/265} and
  \url{https://gitlab.com/ligolang/ligo/-/issues/1248},
data types \url{https://github.com/OCamlPro/liquidity/issues/266}}
That means that one is forced to use primitive container types to write programs.
Therefore, the functions on lists and finite maps must be replaced with ``native'' versions in the extracted code.
We achieve this by providing a translation table that maps names of Coq functions to the corresponding Liquidity/CameLIGO primitives.
Moreover, since the recursive functions can take only a single argument, multiple arguments need to be packed into a tuple.
The same applies to data type constructors since the constructors take a tuple of arguments.
Currently, the packing into tuples is done by the pretty-printers after verifying that constructors are fully applied.

Another issue is related to the type inference in Liquidity and CameLIGO.
Due to the support of overloaded operations on numbers, type inference requires type annotations.
We solve this issue by providing a ``prelude'' for extracted contracts that specifies all required operations on numbers with explicit type annotations.
This also simplifies the remapping of Coq operations to the Liquidity/CameLIGO primitives.
Moreover, we produce type annotations for top-level definitions.

In Coq \icode{Record}s are simply inductive types with one constructor.
At the pretty-printing stage we identify any such types and print them as native records. Liquidity does not allow records with only a single field, or inductives with one constructor.
In this case we print the type as an alias for the type of the field/constructor.
For example, consider the Coq record below, and the function \icode{get\_x} which retrieves the \icode{x} field of the record using Coq's built-in record projection syntax.
\begin{lstlisting}
  Record A := {
    x : nat;
  }.
  Definition get_x (n : A) : nat := n.(x).
\end{lstlisting}
\noindent%
This printed to Liquidity as
\begin{lstlisting}
  type a = nat
  let get_x (n : a) = n
\end{lstlisting}
\noindent%
Note in particular how the projection \icode{a.(x)} is printed simply as \icode{a}.

As a consequence of these restrictions one Liquidity, one should use either type aliases or single-field records in place of inductives with one constructor in the Coq code.
These restrictions only apply if the contract is to be extracted to Liquidity. For the other target languages these restrictions don't apply.

\paragraph*{Higher-order functions in Liquidity.}
Some standard functional programming patterns do not work in Liquidity due to some non-standard features of its type system.
For example, the type of a closure contains information about the environment to which it is closed.%
\footnote{See \url{https://github.com/OCamlPro/liquidity/issues/264}}
For that reason, some programs, which are completely unproblematic in many functional languages are not accepted by the Liquidity compiler.
For example, the following program refuses to compile
\begin{lstlisting}
let my_map (f : int -> int) (xs : int list) =
  List.map f xs

let bar (i : int) (xs : int list) =
  my_map (fun (x : int) -> x + i) xs
\end{lstlisting}%
producing a type error \texttt{Types ((int -> int)[@closure :int]) and int -> int are not compatible}.
The \icode{my_map} function expects a function of type \icode{nat -> nat}, but the call of \icode{my_map} in the body of \icode{bar} gets a function of a different type: \icode{(int -> int)[@closure :int]}, where \icode{:int} refers to the type of the variable \icode{i}.
This makes using higher-order functions highly problematic.
Moreover, this problem extends to closures returned from different branches of \icode{match} expressions, limiting the number of programs that one can extract to Liquidity without additional efforts.

\paragraph*{Handling absurd cases}.
We follow the general strategy outlined in~\cref{sec:false-elim}.
Both Liquidity and CameLIGO feature an effectful operation \icode{failwith}, which allows for interrupting the contract execution.
We identify pattern-matchings with no branches at the pretty-printing stage and insert the \icode{failwith} operation.
However, inlining some constants (e.g.\ \icode{False_rect}) is required in order to make examples like \icode{safe_head} to compile with the CameLIGO compiler, otherwise, extraction produces polymorphic definitions, which are not supported.
For Liquidity, the situation is somewhat worse.
The \icode{failwith} works as expected, however, closure types carry information about the environment wrt.\ which the are closed.
Dependent pattern-matching in Coq produces code with many closures, which are not accepted by the Liquidity compiler.
Therefore, extraction of programs to Liquidity that extensively uses dependent pattern-matching is currently limited.

\paragraph*{Explicit type annotations in CameLIGO.}
LIGO's typechecker is not able to infer types in some instances. These are
\begin{itemize}
  \item 0-ary constructors of polymorphic types, e.g.\ \icode{None} and the empty list \icode{[$\,$]};
  \item function types and in particular arguments of lambda expressions;
  \item the type of \icode{failwith}.
\end{itemize}
Therefore we need to add explicit type annotations to these terms.
To do this, we augment \CICbox{} terms with their types, obtained using the erasure procedure for types~\cref{sec:erasure-for-types}.
We designed a general annotation procedure that can add arbitrary data to the $\CICbox$ AST nodes without changing the AST representation itself.
This is achieved using dependent types: we implement a procedure that for each AST node builds a (nested) product type recursively.
The fragment of the procedure is given below.
\begin{lstlisting}
Fixpoint annots {A : Type} (t : term) : Type :=
  match t with
  | tLambda _ body => A * annots body
  | tApp hd arg => A * (annots hd * annots arg)
  ...
  end.
\end{lstlisting}

For the use case of CameLIGO, we specialise our annotation machinery to \icode{box_type} giving us the type of annotated terms \icode{annots box_type t} for any term \icode{t} of $\CICbox$.
This type-augmented representation is (optional) part of our intermediate representation $\IR$.

The CameLIGO pretty-printer function recurses on the annotated terms and utilises the typing information whenever is necessary.
Since the annotation pass is done separately and independently from the pretty-printer, it may be used for other purposes or new target languages in the future.

\paragraph*{No polymorphic types in CameLIGO.} Unlike Liquidity, CameLIGO does not currently support user-defined polymorphic types, but there is ongoing work to support polymorphic types in the near future\footnote{\url{https://gitlab.com/ligolang/ligo/-/merge_requests/1173}}.
One possibility to circumvent this restriction is to implement a full specialisation pass that produces completely monomorphised code.
However, with the prospect of support for polymorphic types, we instead simply ignore this restriction, although we are aware not all the examples will type check currently.

\paragraph*{Integration.}
In order to generate code for a contract's entry points (functions through which one can interact with the contract), we need to wrap the calls to the main functionality of the contract into a \icode{match} construction.
This is required because the signature of the entry point in Liquidity and CameLIGO is \icode{params -> storage -> (operation list)$\ $* storage}, where \icode{params} is a user-defined type of parameters, \icode{storage} a user-defined state and \icode{operation} is a transfer of contract call.
The signature looks like a total function, but since Liquidity and CameLIGO support a side effect \icode{failwith}, the entry-point function can still fail.
On the other hand, in our Coq development, we use the \icode{option} monad to represent computations that can fail.
For that reason, we generate a wrapper that matches on the result of the extracted function and calls \icode{failwith} if it returns \icode{None}.

The \icode{ChainBase} type class represents an address abstraction, specifying which properties are required from a type of addresses for accounts.
Smart contracts defined using te execution layer infrastructure are abstracted over a \icode{ChainBase} instance.
That means that types \icode{Chain} and \icode{ContractCallContext}, along with \icode{init} and \icode{receive} functions will get an additional parameter corresponding to the \icode{ChainBase} instance.
When printing the contract code, we need to remap \icode{Chain} and \icode{ContractCallContext} to their representation in the target language, and the dependency on \icode{ChainBase} makes it problematic.
We define a specialisation procedure that specialises all definitions dependent on \icode{ChainBase} to an axiomatic instance and removes the corresponding parameter.
Currently, this procedure is defined on PCUIC representation and is not verified.

\paragraph*{Examples.}
The extracted counter contract code to Liquidity and CameLIGO is given, respectively, in \cref{fig:extracted-liquidity} and \cref{fig:extracted-cameligo}.
We omit some wrapper code and the ``prelude'' definitions and leave the most relevant parts (see \cref{appendix:counter-liquidity} and \cref{appendix:counter-cameligo} for the full versions).
As one can see, the extraction procedure removes all ``logical'' parts from the original Coq code.
Particularly, the \icode{sig} type of Coq becomes a simple wrapper for a value (\icode{type 'a sig_ = 'a} in the extracted code).
Currently, we resort to an ad hoc remapping of \icode{sig} to the wrapper \icode{sig_} because Liquidity and CameLIGO do not support variant types with only a single constructor.
Ideally, this class of transformations can be added as an optimisation for inductives types with only one constructor taking a single argument.
This example shows that for certain target languages optimisation is a necessity rather than an option.

We show the extracted code for the \icode{coq_inc_counter} function and omit \icode{coq_dec_counter}, which is extracted in a similar manner.
These functions are called from the \icode{counter} function that performs input validation.
Since the only way of interacting with the contract is by calling \icode{counter} it is safe to execute them without additional input validation, exactly as it is specified in the original Coq code.

Apart from the example in \cref{lst:counter}, we successfully applied the developed extraction to several variants of the counter contract, to the crowdfunding contract described in~\citep{ConCert}, the contracts from~\cref{sec:escrow,sec:boardroom} and to an interpreter for a simple expression language.
The latter example shows the possibility of extracting certified interpreters for domain-specific languages such as Marlowe~\citep{Marlowe}, CSL~\citep{Henglein:CSL} and the CL language~\citep{CertFinContr,CertifiedCompCL}.
This represents an important step towards safe smart contract programming.
The examples show that smart contracts fit well to the fragment of Coq that extracts to well-typed Liquidity and CameLIGO programs.
Moreover, in many cases, our optimisation procedure removes all the boxes resulting in cleaner code.

\paragraph*{Gas consumption.}
Running smart contracts on a blockchain requires so-called ``gas'', which serves as a measure of computational efforts.
The execution environment calculates the gas consumption according to the cost of each operation and uses it to terminate the execution if the maximum gas consumption is reached.
The caller pays a fee in the blockchain's internal currency for calling a contract.
If the fee is too low and the gas consumption is too high, there is a chance that the transaction will not be included in a block.
This behaviour is slightly different from Ethereum, but we will not provide the details here.

We have deployed and executed some of our extracted contracts on test networks (these networks use the same execution model as the main one, but no real money is required to run contracts).
Comparing the gas consumption shows that extracted contracts perform well in the realistic setting, even though, the extracted code consumes more gas compared to a hand-written implementation.
We have compared the ERC20 token implementation against the Liquidity code from the online IDE.
The extracted code consumes 2--2.5 times more gas, but the consumption is quite far from reaching the hard limit on a single contract call.
We have also experimented with the prototype DSL interpreter extracted from our Coq developments on the Tezos network.
The recommended fees, converted to US dollars were lower than $\$0.03$ even for DSL programs with 100 instructions.
Such transaction costs can be considered negligible.
Most of the gas consumption can be attributed to type checking of a contract for each call, which, of course, depends on its length and complexity.
However, gas consumption and the associated fees become smaller with each update of the Tezos network, making the transaction fees negligible for many common use cases.
The threshold on gas consumption also increases, allowing for more expressive smart contracts.
Therefore, our smart contract extraction is able to deliver verified code with reasonable execution costs.

\subsection{Extracting to Elm}\label{sec:elm-extraction}
Elm~\citep{ElmInAction} is a general purpose functional language used for web-development.
It is based on an extended variant of the Hindley-Milner type system and can infer types without user-provided type annotations in most situations.
However, we generate type annotations for top-level definitions to make the extracted code more readable.
Moreover, unlike Liquidity, there is no restriction in Elm regarding data types with one constructor.
That allows implementing a simple extraction procedure for Coq records as data types with one constructor and projections defined as functions that pattern-match on this constructor.
Compared to Liquidity and CameLIGO, Elm is a better target for code extraction, since it does not have some limitations pointed out in \cref{sec:liquidity-cameligo-extraction}.

Extraction to Elm also poses some challenges.
For example, Elm does not allow shadowing of variables and definitions.
Since Coq allows for a more flexible approach to naming, one has to track scopes of variables and generate fresh names in order to avoid clashes.
The syntax of Elm is indentation sensitive, so we are required to track indentation levels.
Various naming conventions apply to Elm identifiers, e.g.\ function names start with a lower-case character, types and constructors --- with an upper case character, requiring some names to be changed when printing.

\paragraph*{Handling absurd cases.}
  We follow the general strategy outlined in~\cref{sec:false-elim}.
  Elm is a pure functional language and does not feature exceptions, which we could use to handle the absurd cases.
  However, there is one side-effect at our disposal, namely, non-termination.
  Therefore, we define the following constant.
  \begin{lstlisting}
    false_rec : () -> a
    false_rec _ = false_rec ()
  \end{lstlisting}
  \noindent%
  At the pretty-printing stage, we identify pattern-matchings with no branches and insert a call to \icode{false_rec}.

\paragraph*{Examples \coqref{extraction/examples/ElmExtractExamples.v}{journal-2021/extraction/examples/ElmExtractExamples.v}.}
We tested the extracted code with the Elm compiler by generating a simple test for each extracted function.
We implemented several examples by extracting functions on lists from the standard library of Coq, functions using subset types and functions that eliminate absurd cases by exploiting contradictions.
All our examples resulted in well-typed Elm code after extraction.
Particularly, in~\cref{fig:extracted-safe-head-elm} one can see the \icode{safe\_head} function (a head of a non-empty list) from~\cref{sec:false-elim}.
The example uses the elimination principle \icode{False_rect} in the case of an empty list, exploiting the contradiction with the assumption that the input list is non-empty.
We used the usual style of writing functions with dependent types in Coq with the help of the \icode{Program} tactic.

As a result, the logical parts corresponding to proofs are erased and Coq's implementation of subset types is extracted as a simple wrapper \icode{type Sig a = Exist a}.
In the impossible (absurd) case, \icode{safe\_head} calls \icode{false_rect}, which is implemented in terms of \icode{false_rec}, using our strategy of handling absurd cases.
We also extract a example function that uses \icode{safe\_head}:
\begin{lstlisting}
Program Definition head_of_repeat_plus_one {A} (n : nat) (a : A) : A
   := safe_head (repeat a (1+n)).
\end{lstlisting}
\noindent%
Clearly, it is always safe to take the first element from a list that is generated by repetition $1+n$ times.
Therefore, the whole program is safe to use and the absurd case will never be hit.
When extracting programs as libraries, one could move \emph{some} static checks to runtime to ensure that the invariants, expected by dependently typed functions are preserved.

We also extract the Ackermann function \icode{ackermann : nat * nat -> nat} defined using well-founded recursion which uses the lexicographic ordering on pairs.
This shows that extraction of definitions based on the accessibility predicate \icode{Acc} is possible.
Computation with \icode{Acc} is studied in more detail in~\citep{Equations}.

\lstset{basicstyle=\scriptsize}
\begin{figure*}
  \begin{subfigure}[t]{.47\textwidth}
    \begin{lstlisting}[language=elm]
type List a
  = Nil
  | Cons a (List a)


-- appending two lists
app : List a -> List a -> List a
app l m =
  case l of
    Nil ->
      m
    Cons a l1 ->
      Cons a (app l1 m)


-- reversing a list
rev : List a -> List a
rev l =
  case l of
    Nil ->
      Nil
    Cons x l2 ->
      app (rev l2) (Cons x Nil)
    \end{lstlisting}
    \caption{\icode{app} and \icode{rev} in Elm}\label{fig:extracted-map-elm}
  \end{subfigure}
  \begin{subfigure}[t]{.47\textwidth}
    \begin{lstlisting}[language=elm]
false_rec : () -> a
false_rec _ = false_rec ()

-- definitions of Nat and List are omitted
...

type Sig a = Exist a

proj1_sig : Sig a -> a
proj1_sig e =
  case e of
    Exist a -> a

false_rect : () -> p
false_rect p = false_rec ()

safe_head : Sig (List a) -> a
safe_head non_empty_list =
  (case proj1_sig non_empty_list of
     Nil -> \x -> false_rect ()
     Cons hd tl -> \x -> hd) ()

head_of_repeat_plus_one : Nat -> a -> a
head_of_repeat_plus_one n a =
  safe_head (Exist (repeat a (add (S O) n)))
    \end{lstlisting}
    \caption{\icode{safe\_head} in Elm}\label{fig:extracted-safe-head-elm}
  \end{subfigure}
  \caption{Extracted test code in Elm.}\label{fig:extracted-code-elm}
\end{figure*}

\lstset{basicstyle=\footnotesize}

\paragraph*{Verified Web Application \coqref{extraction/examples/ElmForms.v}{journal-2021/extraction/examples/ElmForms.v}.}
We develop a more Elm-specific example in Coq: a simple web application.
A typical Elm web application follows the Elm architecture (TEA):
\begin{enumerate}
  \item Model --- a data type representing the state of the application.
  \item Update --- a function that takes a message, a previous state (model instance) and returns a new state and, potentially, commands (e.g.\ sending data to a server, etc.)
  \item View --- a function that turns the model into HTML.
    HTML is generated using special Elm functions, available as part of the Elm standard library.
\end{enumerate}
\noindent%
If we look at the first two items, they look very similar to the smart contract execution model.
At the moment, we do not provide an Elm-specific execution model as part of our framework, but we can leverage Coq dependent types to encode some invariants of the model and then use our extraction pipeline to produce an Elm web application.
Therefore, we implement the model and the update functionality along with validation rules in Coq, extract it to Elm and combine with hand-written rendering code (the view).

The example we consider is inspired by the Elm guide on forms.
Our application consists of an input form, a validator and a view for rendering a list of valid items.
The input form consists of three fields: a username, user's password and a field to re-enter the password.
In Elm, we model it with the following code.

\begin{lstlisting}
Record Entry := { name : string;
                  password : string;
                  passwordAgain : string }.
\end{lstlisting}
\noindent%
This part of the model contains ``raw'' data, as entered by a user.
We define then ``valid data'' using the subset types of Coq.

\begin{lstlisting}
Definition ValidEntry :=
  {entry : Entry | entry.(name) $\ne$ "" /\
                   8 <=? String.length entry.(password) /\
                   entry.(password) =? entry.(passwordAgain)}.
\end{lstlisting}
\noindent%
We use the boolean versions of less-or-equal (\icode{<=?}) and equality (\icode{=?}) on strings, which are implicitly coerced to propositions using \icode{is_true (p : bool) : Prop := p = true}.
This representation makes the interaction with the validation function easier.
Then, we define a type of entries that we are going to store in the list of users in the model.
In the same way, we define what it means to be valid for such stored entries.
\begin{lstlisting}
Record StoredEntry :=
  { seName : string; sePassword : string }.

Definition ValidStoredEntry :=
  { entry : StoredEntry | entry.(name) $\ne$ "" /\ entry.(password) =? entry.(passwordAgain) }.
\end{lstlisting}
\noindent%
Having defined \icode{ValidStoredEntry}, we can then proceed with the definition of a model for the whole application.
\begin{lstlisting}
Record Model :=
  { (** A list of valid entries such with unique user names *)
    users : {l : list ValidStoredEntry | NoDup (seNames l)};
    (** A list of errors after validation *)
    errors : list string;
    (** Current user input *)
    currentEntry : Entry }.
\end{lstlisting}
\noindent%
As one can see, users in our model are represented as a list of valid entries without duplication of names.
Next, we define the messages for updating the model.

\begin{lstlisting}
(** Messages for updating the model according to the current user input *)
Inductive Msg :=
  | MsgName (_ : string)
  | MsgPassword (_ : string)
  | MsgPasswordAgain (_ : string).

(* Messages for updating the current entry and adding the current entry
   to the list of users *)
Inductive StorageMsg :=
   Add
 | UpdateEntry (_ : Msg).
\end{lstlisting}
\noindent%
Now, we can define a function that performs updates to the model by interpreting the messages it receives.
\begin{lstlisting}
Program Definition updateModel : StorageMsg -> Model -> Model * Cmd StorageMsg
  := fun msg model =>
       match msg with
       | Add => match validateModel model with
               | [] => let validEntry : ValidEntry := model.(currentEntry) in
                      let newValidStoredEntry : ValidStoredEntry :=
                        toValidStoredEntry validEntry in
                      let newList := newValidStoredEntry :: model.(users) in
                      (model<| users := newList |>, none)
               | errs => (model<| errors := errs |>, none)
                end
       | UpdateEntry entryMsg =>
              (model<|currentEntry := updateEntry entryMsg model.(currentEntry) |>, none)
       end.
\end{lstlisting}
\noindent%
We use the record update notation \icode{model<| users := newList |>} that uses type classes and Template Coq-based generation of field setters (part of our development).
We also use the standard way of working with subset types in Coq using \icode{Program} command that allows writing code in the style of regular functional programming while manipulating richer types under the hood.
\icode{Program} inserts projections from the values of subset types and constructs values, leaving the proof component as an obligation that the user can prove later.

The main idea behind having valid entries is that most of the functionality of our web application manipulates valid data.
This guarantees that no invariants can be broken by these functions.
The validation is performed only once, at the ``entry point'' of our application, the \icode{updateModel} function and it is driven by the validity predicates of the components of the model.
Therefore, when writing a validation function, it would be impossible to miss some validation rule, because valid data requires explicit proofs of validity.
Since the Elm architecture guarantees that the only way our model is updated when users interact with the web application is by calling the \icode{updateModel} function, we know that in the extracted code the model invariant will not be broken.

The term produced by \icode{Program} might be quite complex, due to the transformations and elaboration required to produce a fully typed term.
Our extraction pipeline is able to cope with the terms generated by \icode{Program} and can be run completely in Coq itself.
We define the required remappings to replace the usage of standard functions with Elm counterparts.
E.g.\ we remap Coq types \icode{string}, \icode{list}, \icode{bool} and the product type to the corresponding types in Elm.
We also remap natural numbers of Coq to type of bounded integers \icode{Int}.
In principle, using bounded numbers might be a problem, but in our case, the only use of numbers is for computing the password length, and \icode{String.length} in Elm has type \icode{String -> Int}.
Therefore, our choice is coherent with the assumptions about string length in Elm.

We use the inlining pass in the pipeline to inline some of the record update infrastructure.
Inlining also prevents us from generating a type alias (related to the records update infrastructure) that is invalid in Elm due to an unused type parameter, which is not allowed.

Overall, we show that one can use the usual certified programming style in Coq in order to implement the logic of a web application that can be then extracted to a fully functional Elm web application (provided that the view functionality is written directly in Elm).
The generated application is well-typed in Elm, even though we have used dependent types extensively.

\subsection{Extracting to Rust}\label{sec:rust-extraction}
Rust is a mixed paradigm general-purpose programming language that features many of the same concepts as functional programming languages.
It is aimed at being a fast language with low overhead, which also makes it an attractive smart contract programming language.
Therefore it provides a lot of control and is also a relatively low-level programming language.
The Concordium blockchain toolchain uses Rust as its primary programming language for writing smart contracts.
The actual code that is executed on-chain is WebAssembly.
WebAssebmly is designed to be a safe, portable and efficient low-level language with well-defined semantics making it well-suited for verification in a proof assistant~\citep{Watt:WebAsm}.
Like Rust, WebAssembly does not feature a garbage collector, making it a good target for compiling Rust.

When writing smart contracts in Rust, the implementors have more ways of controlling performance.
One of the most expensive operations on blockchains is updating the contract's state.
Rust allows for destructive updates of the mutable contract state with the precise control of the serialisation/deserialisation process allowing for careful performance tuning.
Using the Concordium toolchain, smart contracts written in Rust are compiled to WebAssembly modules using LLVM.
The WebAssembly modules can be then deployed and executed on-chain.

Rust has a powerful functional subset that includes
\begin{itemize}
    \item Sum/product types
    \item Pattern matching
    \item Higher-order functions and closures
    \item Immutability by default
    \item Everything-is-an-expression
    \item A Hindley-Milner (without let-polymorphism) based type system
\end{itemize}
These features make Rust a suitable and relatively straightforward target for printing from $\IR$.
However, as Rust is a low-level language giving a lot of control, it also comes with its own set of challenges.

\paragraph*{Extracting data types.}
In Rust, the programmer controls whether fields of data structures are stored by-value or through indirection.
For recursive data structures, such as linked lists, it is necessary to use indirection since otherwise, the size of the data type would be infinite.
Concretely, this means that a type such as
\begin{lstlisting}
Inductive list (A : Type) :=
  | nil
  | cons (head : A) (tail : list A).
\end{lstlisting}%
cannot be extracted in a straightforward way where the tail is just of type \icode{list A}.
Instead, it is necessary to use indirection to store a form of pointer to the tail of the list.
In Rust, there are several ways to store indirection, including raw pointers, borrowed references, owned references (the Box type) and through reference counting (the Rc and Arc types).
The benefit of the Box, Rc, and Arc types is that ownership is managed implicitly, while for raw pointers and borrowed references it is necessary to store the data somewhere else.

Since functional languages generally rely on sharing to perform well the same sharing should be supported in the final extracted Rust program.
In particular, this disqualifies owned references as those can only be shared through expensive copying.
Reference counted types in Rust require explicit cloning to manually indicate when the reference count must be incremented.
This complicates extraction significantly as extraction then has to determine that it needs to insert such clonings when passing arguments and when capturing local variables for closures.

As a result of these considerations, the extraction uses borrowed references to store nested data types.
Such references are trivially copyable and can be shared freely, but require that the data be stored somewhere else.
Additionally, this requires data structures to be generalised over a lifetime.
For uniformity, we add a lifetime to all data types we extract.
As Rust datatypes must use all lifetimes and type parameters they introduce, the extraction also adds ``phantom'' uses of these through the use of \icode{PhantomData}, a zero-cost Rust type meant to specify to the Rust compiler that a lifetime or type parameter is allowed to be unused.
For uniformity, such PhantomData types are emitted as the first member of all data types in all constructors, leading to a final extraction of lists as

\begin{lstlisting}[language=rust]
enum list<'a, A> {
  nil(PhantomData<&'a A>),
  cons(PhantomData<&'a A>, A, &'a list<'a, A>)
}
\end{lstlisting}
The question of ownership is treated next.

\paragraph*{Memory model differences.}
Rust is an unmanaged language without a garbage collector.
When extracting from a language like Coq, in which all memory allocation is handled implicitly for the programmer, this leads to some challenges.
This is made significantly easier when it is noted that smart contract execution is self-contained and very short-lived.
Due to this, it is feasible to allocate as much memory as necessary during the execution and then clean up only after the execution is done, a technique known as region-based memory allocation.
The extraction can thus use an off-the-shelf library that implements region-based memory allocation; in particular, the Bumpalo library is used.

For more general-purpose extraction of programs that may be long-running, we are considering using a conservative garbage collector such as the Boehm-Demers-Weiser~\citep{BoehmWeiserGC,BoehmDemersGC} garbage collector.
Here the challenge lies in implementing the right heuristics to figure out when garbage collection should be invoked during an extracted program.

Extraction produces a structure \icode{Program} that contains the region (or arena) of memory that can be allocated from.
The entire program is extracted as methods on this structure that can then access the region when memory allocation is required.
As an example, consider the function \icode{add : nat -> nat -> nat}, which is extracted with all of its dependencies as
\begin{lstlisting}[language=rust]
pub enum nat<'a> {
  O(PhantomData<&'a ()>),
  S(PhantomData<&'a ()>, &'a nat<'a>)
}

struct Program {
  __alloc: bumpalo::Bump,
}

impl<'a> Program {
  fn new() -> Self {
    Program {
      __alloc: bumpalo::Bump::new(),
    }
  }

  fn alloc<T>(&'a self, t: T) -> &'a T {
    self.__alloc.alloc(t)
  }

  fn closure<TArg, TRet>(&'a self, F: impl Fn(TArg) -> TRet + 'a) -> &'a dyn Fn(TArg) -> TRet {
    self.__alloc.alloc(F)
  }

  fn add(&'a self, n: &'a nat<'a>, m: &'a nat<'a>) -> &'a nat<'a> {
    match n {
      &nat::O(_) => { m },
      &nat::S(_, p) => { self.alloc(nat::S(PhantomData, self.add(p, m))) },
    }
  }
  fn add__curried(&'a self) -> &'a dyn Fn(&'a nat<'a>) -> &'a dyn Fn(&'a nat<'a>) -> &'a nat<'a> {
    self.closure(move |n| { self.closure(move |m| { self.add(n, m) }) })
  }
}
\end{lstlisting}
Our Rust extraction also supports remapping and that \icode{nat} normally would be remapped to either a big-integer type or to the \icode{u64} type using checked arithmetic.

\paragraph*{Handling 'monomorphised' closures.}
In order to handle polymorphic functions (functions with type parameters), the Rust compiler performs a transformation called \emph{monomorphisation}.
That means that the compiler generates copies of a generic function with parameters replaced with concrete types, used in the program.
Rust implements closures in an efficient way by combining their code with the environment they capture into an anonymous type.
Functions can be monomorphised with respect to these types, allowing the use of closures to be a zero-cost abstraction.
For example, closures can be inlined as if they are normal functions or stored directly in data structures.
However, the semantics of such closures are different from the semantics of closures in traditional functional languages.

In Coq, a closure behaves like any other function and is fully compatible with other functions of that function type.
For example, it is possible and unproblematic to store multiple different closures of the same function type in a list.
This uniform behaviour does not carry over to Rust's default treatment of closures: when storing a closure in a list, the list must be typed over the anonymous closure type that the compiler has generated automatically.
Therefore, it is not possible to store two different closures, even of the same function type, in such a list.

Rust still allows for semantics that match Coq's at the cost of some performance through \emph{trait objects}.
Trait objects use virtual dispatch to allow for example closures to behave uniformly as functions, hiding away the associated environment.
Trait objects can exist only as a reference and extraction must thus allocate closures and turn them into references.
The extraction automatically provides the helper function \icode{closure} that performs this allocation using the same region-based allocation as described above.
In some cases the Rust compiler requires annotations when using closures through allocated trait objects.
Therefore, we use the following wrapper to aid the type inference.
\begin{lstlisting}[language=rust]
fn hint_app<TArg, TRet>(f: &dyn Fn(TArg) -> TRet) -> &dyn Fn(TArg) -> TRet {
  f
}
\end{lstlisting}
\noindent%
We insert the wrapper whenever we have an application of a closure.

\paragraph*{Partial applications.}
Rust requires all functions to be fully applied when called, unlike Coq which supports partial application.
Partial applications can be emulated easily through closures, by generating both curried versions and uncurried versions of functions.
However, using closures is less efficient, so as an optimization the extraction avoids closures when possible.
Concretely, this results in both curried and uncurried versions as is seen in the extraction above, with the curried version calling into the uncurried version.

\paragraph*{Internal fixpoints.}
Coq supports recursive closures through the \icode{fix} construct.
In comparison, Rust does not have similar support for recursive closures and supports only recursive local functions which do not allow capturing.
This means that only top-level recursive Coq functions can straightforwardly be made recursive during extraction; when a fixpoint is used internally (for example, through a \icode{let fix} binding), there is no simple way to extract this.
To work around this issue, we apply a technique known as ``Landin's knot''~\citep{Landin:TheME}.
Namely, our extraction uses recursion through the heap.
Concretely, when an internal fixpoint is encountered, extraction produces code that allocates a cell on the heap to store a reference to the closure.
The closure can access this heap cell and thus access itself when it needs to recurse.
To exemplify, a straightforward definition of the Ackermann function in Coq uses nested recursion:
\begin{lstlisting}
Fixpoint ack (n m : nat) : nat :=
  match n with
  | O => S m
  | S p => let fix ackn (m : nat) :=
               match m with
               | O => ack p 1
               | S q => ack p (ackn q)
               end
           in ackn m
  end.
\end{lstlisting}
and extraction produces
\begin{lstlisting}[language=rust]
fn ack(&'a self, n: &'a Nat<'a>, m: &'a Nat<'a>) -> &'a Nat<'a> {
  match n {
    &Nat::O(_) => { self.alloc(Nat::S(PhantomData, m)) },
    &Nat::S(_, p) => {
      let ackn = {
        let ackn = self.alloc(std::cell::Cell::new(None));
        ackn.set(Some(
          self.closure(move |m2| {
            match m2 {
              &Nat::O(_) => {
                self.ack(
                  p,
                  self.alloc(Nat::S(PhantomData, self.alloc(Nat::O(PhantomData))))
                )
              },
              &Nat::S(_, q) => { self.ack(p, ackn.get().unwrap()(q)) },
            }
          })));
        ackn.get().unwrap()
      };
      ackn(m)
    },
  }
}
\end{lstlisting}

\paragraph*{Handling absurd cases.}
  We follow the general strategy outlined in~\cref{sec:false-elim}.
  The natural choice for implementing the elimination principle for an empty type is to use Rust's \icode{panic!} macro.
  In this case the elimination principle for \icode{False} extracts to the following Rust code.
  \begin{lstlisting}[language=rust]
fn False_rect<P: Copy>(&'a self, u: ()) -> P {
  panic!("Absurd case!")
}
\end{lstlisting}
  We identify pattern-matchings with no branches at the pretty-printing stage and insert the \icode{panic!} macro.

\paragraph*{Integrating with Concordium \coqref{extraction/theories/ConcordiumExtract.v}{journal-2021/extraction/theories/ConcordiumExtract.v}.\newline}
Concordium maintains the smart contract state in a serialized form, i.e. as an array of bytes.
Similarly, when a smart contract is called, its message is passed as an array of bytes.
To aid in conversion between the smart contract's data types and these byte arrays, the Concordium toolchain provides automatic derivation of serializers and deserializers between arrays of bytes and standard Rust data types.
This conversion, however, does not support references, as it is unclear how to deserialize into a reference.
In addition, the ConCert smart contracts extracted are not immediately compatible with the signatures expected by Concordium.

To aid in adapting between ConCert and Concordium a standard library is provided by ConCert.
This standard library includes several helper types that extracted smart contracts depend on, and additionally also provide procedural macros that can be used to derive serializers and deserializers that, through the use of regions, support deserializing references.
When a smart contract is extracted, it automatically has serializers and deserializers derived for its structures, and the extraction takes care to generate glue code that properly performs deserialization and serialization with a proper region.
Finally, the glue code also adapt between Concordium's expected smart contract signature and the one extracted by ConCert.\\
We proceed to highlight some case studies using the ConCert framework.

\section{The Escrow Contract}\label{sec:escrow}
As an example of a nontrivial contract, we can extract we describe in this section an \textit{escrow} contract~\coqref{execution/examples/Escrow.v}{journal-2021/execution/examples/Escrow.v}.
The purpose of this contract is to enable a seller to sell goods in a trustless setting via the blockchain.
The Escrow contract is suited for goods that cannot be delivered digitally over the blockchain; for goods that can be delivered digitally, there are contracts with better properties, such as FairSwap~\citep{FairSwap}.

Because goods are not delivered on-chain there is no way for the contract to verify that the buyer has received the item.
Instead, the contract incentivises the parties to follow the protocol by requiring that both parties commit additional money that they are paid back at the end.
Assuming a seller wants to sell a physical item for $x$ amount of currency, the contract proceeds in the following steps:
\begin{enumerate}
\item The seller deploys the contract and commits (by including with the deployment) $2x$.
\item The buyer commits $2x$ before a deadline.
\item The seller delivers the goods (outside of the smart contract).
\item The buyer confirms (by sending a message to the smart contract) that they have received the item.
      They can then withdraw $x$ from the contract while the seller can withdraw $3x$ from the contract.
\end{enumerate}

If there is no buyer who commits funds the seller can withdraw their money back after the deadline.
Note that when the buyer has received the item, they can choose not to notify the smart contract that this has happened.
In this case, they will lose out on $x$, but the seller will lose out on $3x$.
In our work, we assume that this does not happen, and we consider the exact game-theoretic analysis of the protocol to be out of scope.
Instead, we focus on proving the \textit{logic} of the smart contract correct under the assumption that both parties follow the protocol to completion.
The logic of the Escrow is implemented in approx.\ a hundred lines of Gallina code.
The interface to the Escrow is its message type given below.
\begin{lstlisting}
Inductive Msg := commit_money | confirm_item_received | withdraw.
\end{lstlisting}
To state correctness, we first need a definition of what the escrow's effect on a party's balance has been.
\begin{coqdefinition}{\coqref{execution/examples/Escrow.v:net\_balance\_effect}{5eb12f34fc29035eb0909ffdda523c4b70eaa462/execution/examples/Escrow.v\#L537}}[Net balance effect]
Let $\pi$ be an execution trace and $a$ be an address of some party.
Let $T_\text{from}$ be the set of transactions from the Escrow to $a$ in $\pi$, and let $T_\text{to}$ be the set of transactions from $a$ to the contract in $\pi$.
Then the net balance effect of the Escrow on $a$ is defined to be the sum of amounts in $T_\text{from}$, minus the sum of amounts in $T_\text{to}$.
\end{coqdefinition}
\noindent The Escrow keeps track of when both the buyer and seller have withdrawn their money, after which it marks the sale as completed.
This is what we use to state correctness.
\begin{coqtheorem}{\coqref{execution/examples/Escrow.v:escrow\_correct}{5eb12f34fc29035eb0909ffdda523c4b70eaa462/execution/examples/Escrow.v\#L554}}[Escrow correctness]\label{thm:escrow-correct}
  Let $\pi$ be an execution trace with a finished Escrow for an item of value $x$.
  Let $S$ be the address of the seller and $B$ the address of the buyer.
  Then:
  \begin{itemize}
    \item If $B$ sent a \icode{confirm_item_received} message to the Escrow, the net balance effect on the buyer is $-x$ and the net balance effect on the seller is $x$.
    \item Otherwise, the net balance effects on the buyer and seller are both $0$.
  \end{itemize}
\end{coqtheorem}
\noindent%
Below, we show how the informal statement of \cref{thm:escrow-correct} is implemented in Coq using the infrastructure provided by the execution layer (see \cref{sec:ConCert}).
In the comments, we point out the corresponding parts and notations from the informal statement of the theorem.
\begin{lstlisting}
    Theorem escrow_correct
            {ChainBuilder : ChainBuilderType}
            prev new header acts :
    (* For a trace ($\pi$) ending with a successful addition of a block (reachability) *)
    builder_add_block prev header acts = Ok new ->
    let trace := builder_trace new in
    forall caddr,
      env_contracts new caddr = Some (Escrow.contract : WeakContract) ->
      exists (depinfo : DeploymentInfo Setup)
             (cstate : State)
             (inc_calls : list (ContractCallInfo Msg)),
        deployment_info Setup trace caddr = Some depinfo /\
        contract_state new caddr = Some cstate /\
        incoming_calls Msg trace caddr = Some inc_calls /\
        (* the value of the item ($x$) *)
        let item_worth := deployment_amount depinfo / 2 in
        (* the address of the seller $S$ *)
        let seller := deployment_from depinfo in
        (* the address of the buyer $B$ *)
        let buyer := setup_buyer (deployment_setup depinfo) in
        is_escrow_finished cstate = true ->
        (* the net balance effect is $x$ on the seller and $-x$ on the buyer *)
        (buyer_confirmed inc_calls buyer = true /\
         net_balance_effect trace caddr seller = item_worth /\
         net_balance_effect trace caddr buyer = -item_worth \/
         (* otherwise, the net balance effects on the buyer and seller are both $0$. *)
         buyer_confirmed inc_calls buyer = false /\
         net_balance_effect trace caddr seller = 0 /\
         net_balance_effect trace caddr buyer = 0).
\end{lstlisting}
\noindent%
\sloppy%
In Coq, we first prove a slightly more general statement of the theorem (\icode{escrow_correct_strong}), which is then used to prove the statement that corresponds to \cref{thm:escrow-correct}.
The proof is by induction on the structure of the contract's execution trace \icode{ChainTrace}.
We use a specialised induction principle that allows for better proof structure.
Moreover, we provide textual hints for the user for each case when applying the inductive principle in the interactive mode.

\paragraph*{Extracting the contract \coqref{extraction/examples/EscrowExtract.v}{journal-2021/extraction/examples/EscrowExtract.v}.}
We have successfully extracted the escrow contract to Rust, CameLIGO, and Liquidity.
For CameLIGO and Liquidity, we remap the \icode{Amount} type (which is just an alias for \icode{Z}) to \icode{tez}, the on-chain currency.
We also remap the fields of \icode{Chain} and \icode{ContractCallContext} to equivalent API calls in CameLIGO/Liquidity.
For example, the \icode{ctx\_contract\_balance} field of \icode{ContractCallContext} is remapped to \icode{Tezos.balance} for CameLIGO, and \icode{Current.balance} for Liquidity.

Liquidity has a small caveat that it does not allow external functional calls in the initialisation function.
Using the inlining transformation described in \cref{sec:proof-generating}, we ensure that the necessary function definitions are inlined in the initialisation function.
Furthermore, we also inline various monad instances implicitly used in the contract code, such as the instance for \icode{Monad option}, since higher-kinded types are not supported in CameLIGO and Liquidity.

The Rust version of the escrow contract was successfully deployed and instantiated on the Concordium's test network.
This demonstrates that the integration infrastructure is fully functional.
The size of the resulting Wasm executable that was obtained by compiling the extracted contract is about 39KB, while the threshold is 64KB.

\section{The Boardroom Voting Contract}\label{sec:boardroom}
Hao, Ryan and Zieli\'{n}sky developed the Open Vote Network protocol~\citep{open-vote-network}, an e-voting protocol that allows a small number of parties (`a boardroom') to vote anonymously on a topic.
Their protocol allows tallying the vote while still maintaining maximum voter privacy, meaning that each vote is kept private unless all other parties collude.
Each party proves in zero-knowledge to all other parties that they are following the protocol correctly and that their votes are well-formed.

This protocol was implemented as an Ethereum smart contract by McCorry, Shahandashti and Hao~\citep{ethereum-boardroom}.
In their implementation, the smart contract serves as the orchestrator of the vote by verifying the zero-knowledge proofs and computing the final tally.

We implement a similar contract in the ConCert framework~\coqref{execution/examples/BoardroomVoting.v}{journal-2021/execution/examples/BoardroomVoting.v}.
The original protocol works in three steps.
First, there is a sign-up step where each party submits a public key and a zero-knowledge proof that they know the corresponding private key.
After this, each party publishes a commitment to their upcoming vote.
Finally, each party submits a computation representing their vote, but from which it is computationally intractable to obtain their actual private vote.
Together with the vote, they also submit a zero-knowledge proof that this value is well-formed, i.e.\ it was computed from their private key and a private vote (either `for` or `against`).
After all, parties have submitted their public votes, the contract is able to tally the final result.
For more details, see the original paper~\citep{open-vote-network}.
The contract accepts messages given by the type:
\begin{lstlisting}
Inductive Msg :=
| signup (pk : A) (proof : A * Z)
| commit_to_vote (hash : positive)
| submit_vote (v : A) (proof : VoteProof)
| tally_votes.
\end{lstlisting}%
Here, \icode{A} is an element in an arbitrary finite field, \icode{Z} is the type of integers and \icode{positive} can be viewed as the type of finite bit strings.
Since the tallying and the zero-knowledge proofs are based on finite field arithmetic we develop some required theory about $\mathbb{Z}_p$ including Fermat's theorem and the extended Euclidean algorithm.
This allows us to instantiate the boardroom voting contract with $\mathbb{Z}_p$ and test it inside Coq using ConCert's executable specification.
To make this efficient, we use the Bignums library of Coq to implement operations inside $\mathbb{Z}_p$ efficiently.

The contract provides three functions \icode{make_signup_msg}, \icode{make_commit_msg} and \icode{make_vote_msg} meant to be used off-chain by each party to create the messages that should be sent to the contract.
As input, these functions take the party's private data, such as private keys and the private vote, and produces a message containing derived keys and derived votes that can be made public, and also zero-knowledge proofs about these.
We prove the zero-knowledge proofs attached will be verified correctly by the contract when these functions are used.
Note that, due to this verification done by the contract, the contract is able to detect if a party misbehaves.
However, we do not prove formally that incorrect proofs do not verify since this is a probabilistic statement better suited for tools like EasyCrypt or SSProve~\citep{SSProve}.

When creating a vote message using \icode{make_vote_msg} the function is given as input the private vote: either `for`, represented as $1$, and `against`, represented as $0$.
We prove that the contract tallies the vote correctly assuming that the functions provided by the boardroom voting contract are used.
Note that the contract accepts the \icode{tally_votes} message only when it has received votes from all public parties, and as a result stores the computed tally in its internal state.
We give here a simplified version of the full correctness statement which can be found in the attached artifact.
\begin{coqtheorem}{\coqref{execution/examples/BoardroomVoting.v:boardroom\_voting\_correct}{5eb12f34fc29035eb0909ffdda523c4b70eaa462/execution/examples/BoardroomVoting.v\#L814}}[Boardroom voting correct]\label{thm:boardroomvoting-correct}
  Let $\pi$ be an execution trace with a boardroom voting contract.
  Assume that all messages to the Boardroom Voting contract in $\pi$ were created using the functions described above. Then:
  \begin{itemize}
    \item\sloppy If the boardroom voting contract has accepted a \icode{tally_votes} message, the tally stored by the contract equals the sum of private votes.
    \item Otherwise, no tally is stored by the contract.
  \end{itemize}
\end{coqtheorem}
\noindent%
Below, we show how the informal statement of \cref{thm:boardroomvoting-correct} is implemented in Coq using the infrastructure provided by the execution layer (see \cref{sec:ConCert}).
In the comments, we point out the corresponding parts and notations from the informal statement of the theorem.
\begin{lstlisting}
  Theorem boardroom_voting_correct
        (bstate : ChainState)
        (caddr : Address)
        (* For any trace ($\pi$) from the initial state to a reachable state [bstate]  *)
        (trace : ChainTrace empty_state bstate)
        (* a list of all public keys, in the order of signups *)
        (pks : list A)
        (* a function mapping a party to information about them *)
        (parties : Address -> SecretVoterInfo) :
    env_contracts bstate caddr = Some (boardroom_voting : WeakContract) ->
    exists (cstate : State)
           (depinfo : DeploymentInfo Setup)
           (inc_calls : list (ContractCallInfo Msg)),
      deployment_info Setup trace caddr = Some depinfo /\
      contract_state bstate caddr = Some cstate /\
      incoming_calls Msg trace caddr = Some inc_calls /\

      (* assuming that the message sent were created with the functions
         provided by this smart contract *)
      MsgAssumption pks parties inc_calls ->

      (* ..and that people signed up in the order given by 'index' and 'pks' *)
      SignupOrderAssumption pks parties inc_calls ->

      (* ..and that the correct number of people register *)
      (finish_registration_by (setup cstate) < Blockchain.current_slot bstate ->
       length pks = length (signups inc_calls)) ->

      (* then if we have not tallied yet, the tally is none *)
      ((has_tallied inc_calls = false -> tally cstate = None) /\
       (* or if we have tallied yet, the tally is correct *)
       (has_tallied inc_calls = true ->
        tally cstate = Some (sumnat (fun party => if svi_sv (parties party) then 1 else 0)
                                    (map fst (signups inc_calls))))).
\end{lstlisting}

Similarly to the escrow contract from~\cref{sec:escrow}, we first prove a more general theorem using the specialised induction principle for the execution traces.

\paragraph*{Extracting the contract \coqref{extraction/examples/BoardroomVotingExtractionCameLIGO.v}{journal-2021/extraction/examples/BoardroomVotingExtractionCameLIGO.v}.}
The boardroom voting contract gives a good benchmark for our extraction as it relies on some expensive computations.
It drives our efforts to cover more practical cases, and we have successfully extracted it to CameLIGO.

The main problem with extraction for this contract is the use of higher-kinded types.
In particular, the implementation of the contract uses finite maps from the std++ library, which implicitly rely on higher-kinded types.
In addition, the contract uses monadic binds, implemented via type classes that require passing type families around.
Furthermore, the arithmetic operations and developed theory is captured in the type class \icode{BoardroomAxioms (A : Type)}, where \icode{A} is the element type of the finite field, and is instantiated to $\mathbb{Z}_p$ for extraction.
All of this is not representable in prenex-polymorphic type systems, and our target languages follow a similar typing discipline to prenex-polymorphism.
While we could adjust the implementation to avoid relying on higher kinded types, we instead prefer to improve the extraction to work on more examples.
In particular, for our cases, we have identified that a few steps of reduction is enough for most of the higher kinded types to disappear.
For example, the signature of \icode{bind} is \icode{forall m : Type -> Type, Monad m -> forall t u : Type, m t -> (t -> m u) -> m u} which, when it appears in the contract, typically looks like \icode{bind option option_monad ...} where \icode{option_monad} is some constant that builds a record describing the option monad.
After very few steps of reduction, this reduces to the well-known bind for options, which is unproblematic to extract.
At this point, the pre-processing pass (see~\cref{sec:proof-generating}) comes in handy and the inlining functionality is sufficient to produce definitions that are well-typed after extraction.

For the \icode{BoardroomAxioms} type class, on which the entire contract is parameterised over, we would need a specialisation pass similar to the \icode{ChainBase} specialisation described in~\cref{sec:liquidity-cameligo-extraction}.
It could be possible with a more general technique, such as partial evaluation.
We leave this as future work, and in the meantime create a copy of the contract where we have inlined $\mathbb{Z}_p$ in place of \icode{A}~\coqref{execution/examples/BoardroomVotingZ.v}{journal-2021/execution/examples/BoardroomVotingZ.v}.

\section{Related Work}\label{sec:related-work}
\paragraph*{Extraction to statically typed languages.}
The works in this direction are the most relevant for the present development.
By \emph{extraction} we mean obtaining source code in a target language which is accepted by the target language compiler and can be further integrated with existing systems.
Several proof assistants share this feature (Coq~\citep{CoqNewExtraction}, Isabelle~\citep{Isabelle-ExecutingHOL}, Agda~\citep{Kusee:agda-haskell}) and allow targeting conventional functional languages such as Haskell, OCaml or Standard ML.
However, extraction in Isabelle/HOL~\citep{Isabelle-ExecutingHOL} is slightly different from Coq and Agda, since in the higher-order logic of Isabelle/HOL programs are represented as equations and the job of the extraction mechanism is to turn them into executable programs.
Moreover, Isabelle/HOL does not feature dependent types, therefore the type system of programs is closer to the extraction targets, in contrast to Coq and Agda, where one has to make additional efforts to remove proofs from terms.

Clearly, the correctness of the extraction code is crucial for producing correct executable programs.
This is addressed by several developments for Isabelle~\citep{Isabelle-CodeGen,VerfiedCompIsabelle}.
The work~\citep{VerfiedCompIsabelle} features verified compilation from Isabelle/HOL to CakeML~\citep{CakeML}.
It also implements meta-programming facilities for quoting Isabelle/HOL terms similar to MetaCoq.
Moreover, the quoting procedure produces a proof that the quoted terms correspond to the original ones.
The current extraction implemented in the Coq proof assistant is not verified.
Although the theoretical basis for it is well-developed by Letouzey~\citep{letouzey04}, Coq's extraction also includes unverified optimisations that are done together with extraction, making it harder to compare it with the formal treatment given by Letouzey.
So, the unverified extraction even lacks a full paper proof.
Our separation between erasure and optimisation facilitates such comparisons, and allows reuse of the optimisation pass in a standalone fashion in other projects.
The MetaCoq project~\citep{CertErasure} aims to formalise the meta-theory of the calculus of inductive constructions and features a verified erasure procedure that forms the basis for extraction presented in this work.
We also emphasise that the previous works on extraction targeted conventional functional languages (e.g.\ Haskell, OCaml, etc.), while we target the more diverse field of functional smart contract languages.

The authors of~\citep{SinkarovsCockx:ChoosingIsLosing} present an approach for defining embeddings and extraction procedures at the same time in Agda.
The approach is best suited for domain-specific languages and characterises the subset of Agda from which extraction is possible by the successful execution of the extraction procedure.
Currently, it seems impossible to establish semantic preservation properties for the extraction/embedding procedures, because the meta-theory of Agda is not formalised.
In our setting, we mostly work with general-purpose languages.
In this case, applying this approach seems to be problematic, since embedding a general-purpose language can be a non-trivial effort.
However, for certain domain-specific contract languages (e.g.\ Marlowe~\citep{Marlowe}, CSL~\citep{Henglein:CSL}, CL~\citep{CertFinContr,CertifiedCompCL}) the approach of~\citep{SinkarovsCockx:ChoosingIsLosing} looks promising.
It would be interesting to reproduce the approach in Coq, with the additional benefit of reasoning about the semantics preservation using the MetaCoq formalisation.
Currently, we have an example of a simple DSL interpreter extracted from Coq (see examples in~\cref{sec:liquidity-cameligo-extraction}) which could be accompanied by an embedding.

The recent developments in quantitative type theory (QTT)~\citep{Atkey:QTT} offer an interesting perspective on erasure.
QTT allows for tracking the resource usage in types, and this information can be used to identify what can be erased at run-time.
Agda's GHC backend uses QTT-inspired erasure annotations~\citep{Danielsson:AgdaErasure} in order to remove computationally-irrelevant parts of extracted Haskell programs.
However, in our case, it would require significantly changing the underlying theory of Coq.
Therefore, such techniques are currently not available to us.

The implicit calculus of constructions (ICC)~\citep{Miquel:ICC} offers an alternative to using \icode{Prop} for separating the computational content from specifications and proofs.
ICC adds an \emph{implicit product} type $\forall x : T. U$ allowing for quantifying over $x$ without introducing extra binders at the term level.
However, the type checking in ICC is undecidable.
The authors of~\citep{BarrasBernardo:ICCCoq} present an annotated variant ICC$^{*}$, which recovers the decidability.
The terms of ICC$^{*}$ can be extracted to ICC by removing the annotations.
In the PhD thesis by Bernardo~\citep{bernardo:PhD}, ICC and its annotated variant were extended with dependent pairs ($\Sigma$-types), including an implicit version (similarly to the implicit product).
One benefit of using ICC-like type systems is that it allows for more definitional equalities.
E.g.\ for two dependent pairs \icode{(a,p1)} and \icode{a,p2} (\icode{p1} and \icode{p2} are proofs of some property on \icode{a} and \icode{b}) are equal whenever the first components are equal.
The proofs of the same property are definitionally equal in such systems.
The same definitional equality can be obtained in Coq using the universe of definitionally proof-irrelevant propositions \icode{SProp}~\citep{Gilbert:SProp}.
However, ICC$^{*}$ allows for making binders of an arbitrary type irrelevant, prohibiting their use in computationally relevant parts of a program.
Effectively, it means that irrelevant arguments do not occur in terms of pure ICC (after erasure), but can be used without any restrictions in the codomain of the implicit product type.
E.g.\ \icode{fun \{n\} (v : vec n) => n } is ill-typed in ICC$^{*}$, where \icode{vec} is the type of sized lists (also called vectors).
This restriction cannot be expressed using \icode{SProp}.
Moreover, implementing the conversion test through extraction to pure ICC gives a very expressive subtyping relation.
For example, vectors in this system would be subtypes of lists (using the impredicative encodings for vectors and lists).
The approach of ICC looks promising and authors of~\citep{BarrasBernardo:ICCCoq} report that ICC$^{*}$ allows for a simpler implementation.\footnote{A prototype implementation is available for older versions of Coq: \url{http://www.lix.polytechnique.fr/Labo/Bruno.Barras/coq-implicit/}}
However, it seems that ICC has not been extended to handle the full calculus of inductive constructions.

The authors of~\citep{MishraSheard:ErasurePTS} consider an approach similar to ICC in the context of pure type systems (PTS).
The present two calculi: EPTS (Erasure Pure Type Systems --- a calculus of annotated terms, similar to ICC$^{*}$) and IPTS (Implicit Pure Type Systems, similar to ICC).
The EPTS calculus features \emph{phase distinction} annotations for distinguishing between compile-time and run-time computations.
The authors define an erasure procedure from EPTS to IPTS and briefly discuss some implementation issues.
It seems that the implementation of the presented system is not currently available.

\paragraph*{Execution of dependently typed languages.}
Related works in this category are concerned with compiling a dependently-typed language to a low-level representation.
Although the techniques used in these approaches are similar to extraction, one does not need to fit the extracted code into the type system of a target language and is free to choose an intermediate compiler representation.
The dependently typed programming language Idris uses erasure techniques for efficient execution~\citep{Ind-falimilies-indices:Brady}.
The Idris 2 implementation~\citep{Brady:Idris2QTT} implements QTT for both tracking the resource consumption and the run-time irrelevance information.

For the Coq proof assistant, the work~\citep{CIC-type-decorations:Barras2005} develops an approach for efficient convertibility testing of untyped terms acquired from fully typed CIC terms.
The {\OE}uf project~\citep{Oeuf} features verified compilation of a restricted subset of Coq's functional language Gallina (no pattern-matching, no user-defined inductive types --- only eliminators for particular inductives).
In~\citep{ExtractionFiat}, the authors report on the extraction of domain-specific languages embedded into Gallina to an imperative intermediate language that can be compiled to efficient low-level code.
And finally, the certified compilation approach to executing Coq programs is under development in the CertiCoq project~\citep{Anand:CertiCoq}.
The project uses MetaCoq for quotation functionality and uses the verified erasure as the first stage.
After several intermediate stages, C light code is emitted and later compiled for a target machine using the CompCert certified compiler~\citep{2006-Leroy-compcert}.
Since we implement our pass as a standalone optimisation on the same AST that is used in CertiCoq, our pass can be integrated in a relatively straightforward fashion in CertiCoq (see~\cref{sec:conclusions}).

\paragraph*{Formalisation of target languages.}
Another group of works that can be useful for extending the guarantees provided by extraction is the formalisation of the semantics of target languages.
That is, one can add a translation step from $\IR$ to the target language syntax and prove the translation correct.

Ongoing work at Tezos on formalising the semantics of LIGO languages\footnote{\url{https://gitlab.com/ligolang/ligo/-/tree/dev/src/coq}} would allow for connecting our $\IR$ semantics with the CameLIGO semantics, and eventually get a verified pipeline producing verified Michelson code (which is directly executed by the Tezos infrastructure).
Projects like RustBelt~\citep{RustBelt} and Oxide~\citep{RustOxide} are aiming to give formal semantics to Rust.
However, currently, they do not formalise the Rust surface language.

\paragraph*{Dead arguments elimination.}
The techniques of removing computationally useless (dead) code were extensively studied in the context of simply-typed~\citep{Berardi:PruningSTLC} and polymorphic~\citep{Boerio:PruningSF2} $\lambda$-calculi.
The techniques were extended to the calculus of constructions (CoC) by Prost~\citep{Prost:Marking}.
These techniques analyse the terms to identify unused parts and mark them.
As a result, one obtains a \emph{typed} term with some redundancy removed.
This captures the proofs that do not contribute to the final result.

We follow the approach initially developed by Paulin-Mohring~\citep{Paulin-Mohring:Extraction} for CoC and later adapted and extended to CIC by Letouzey~\citep{letouzey04}.
Namely, all computationally irrelevant propositions must be placed in a special universe \icode{Prop}.
However, we apply a pass that removes dead arguments \emph{after} erasure.
Letouzey mentioned in his thesis, that doing so has the benefit of working with smaller terms (since large parts are replaced with the $\Box$ constructor).
Moreover, in~\citep{CoqNewExtraction} he says that implementation of extraction contains ``a workaround designed to safely remove most of the external dummy lambdas''.
We demonstrate that this workaround can be replaced with a more general and principled optimisation (see~\cref{sec:optimisations}).

\section{Conclusion and Future Work}\label{sec:conclusions}
We have presented an extraction pipeline implemented completely in the Coq proof assistant.
This approach has an important advantage: we can use Coq for providing strong correctness guarantees for the extraction process itself by verifying the passes of the pipeline.
The whole range of certified programming and proof techniques becomes applicable since the pipeline consists of ordinary Coq definitions.
Our extraction relies on the MetaCoq verified erasure procedure~\citep{CertErasure}, which we extend with data structures required for extraction to our target languages.
Our pipeline addresses new challenges originating from the target languages we have considered and can be extended with new transformations if required.

The developed approach allows for targeting various functional languages.
Currently, we support two target languages for smart contract extraction (Liquidity and CameLIGO) and two general-purpose languages (Elm and Rust).
Rust is also used as a smart contract language for the Concordium blockchain.
Our experience shows that the extraction is well-suited for Coq programs in a fragment of Gallina that corresponds to a generic polymorphic functional language extended with subset types.
This fragment is sufficient to cover most of the features found in functional smart contract languages and is suitable for extracting many programs to Rust and Elm, resulting in well-typed code.
In general, our pipeline allows for implementing, testing, verifying and extracting programs from Coq to new target languages, while retaining a small TCB.

We have tested our extraction pipeline on various example programs.
In the domain of smart contracts, we have considered several examples both designed for demonstration purposes and representing real-world use cases.
The short descriptions of the contracts are given below.
\begin{itemize}
  \item \icode{Counter} --- a simple contract featuring the increment and decrement functionality (similar to the example in~\cref{lst:counter}, but without using the advanced Coq types).
  \item \icode{Counter (subset types)} --- the example in~\cref{lst:counter}.
  \item \icode{ERC20 token} --- an implementation of a widely used token standard.
  \item \icode{Crowdfunding} --- a smart contract representing a common use case, also knows as Crowdsale, Kickstarter-like contract, ICO contract, etc
  \item \icode{DSL Interpreter} --- a simple interpreter, demonstrating a feasibility of embedding interpreted DSLs.
  \item \icode{Escrow} --- an implementation of an escrow (see~\cref{sec:escrow}).
  \item \icode{Boardroom voting} --- an implementation an anonymous e-voting protocol (see~\cref{sec:boardroom}).
\end{itemize}%
The examples we have considered confirm that our pipeline is suitable for extracting real-world smart contracts.

Our verified optimisation pass is generic, making it applicable for other projects that use MetaCoq erasure.
We have integrated our pass with the CertiCoq project~\citep{Anand:CertiCoq} and have sent a pull request to the CertiCoq repository.%
\footnote{https://github.com/CertiCoq/certicoq/pull/29}
With small modifications, the pass seems to be beneficial for the CertiCoq pipeline and can potentially replace a similar unverified pass, but it is not yet merged into the main CertiCoq development.
The main reason for that is that our optimisation assumes that constants and constructors are applied to all arguments we remove.
That means that there should be an $\eta$-expansion pass in the pipeline, which we solve with our proof generating approach.
However, there is no such pass in CertiCoq, but there are plans to add such a pass to the MetaCoq development.
After that, our optimisation pass could be fully integrated into CertiCoq.

As future work one can imagine various additional and improved optimisations, that fit well with the infrastructure we have developed.
For example, removing singleton inductives (e.g.\ \icode{Acc}), ``unboxing'' the values built from one-argument constructors application (originating from inductive types with one constructor, e.g.\ constructors of a subset type \icode{sig}).
The proof-generating pass allows for inlining and specialising some definitions which might not be typable after extraction, since our targets do not feature unsafe type casts, like OCaml's \icode{Obj.magic}.
Our pipeline is well-suited for adding new conversion-preserving transformations at a very low cost: one just has to write a function, with the signature \icode{global_env -> Result global_env string} and include it in the list of transformations.
The proofs of correctness will be generated automatically after all the transformations have been applied.
For example, the pipeline can accommodate techniques such as partial evaluation as it is presented by~\cite{Tanaka:CoqToC}, which could be implemented in Coq directly (instead of a plugin) using the meta-programming facilities of MetaCoq.

We plan also to improve the boardroom voting contract extraction.
First, we would like to implement more machinery for program specialisation (like partial evaluation mentioned above), making the manual adjustments of the boardroom voting contract unnecessary.
Second, we would like to integrate it with extracted high-performance cryptographic primitives using the approach of FiatCrypto~\citep{FiatCrypto}.
For example, the Open Vote Network protocol, on which the contact is based, depends on computations in a Schnorr group, a large prime-order subgroup of the multiplicative group of integers modulo some prime $p$.
An efficient Rust implementation of a Schnorr group can be obtained from FiatCrypto.
We can then replace our naive implementation by this highly optimized implementation.

Since we have already considered new target languages from the ML-family (Elm, Liquidity and CameLIGO), we expect that our pipeline can be also used for extracting to OCaml, similarly to the standard Coq extraction.
Currently, the standard extraction of Coq implements more optimisations than we support in our pipeline.
However, our development enables adding more verified optimisations in a compositional manner, giving a systematic way of improving the extraction output.
Inserting unsafe type coercions (\icode{Obj.magic}) is currently not supported by our development, due to the absence of such mechanisms in most of our targets.
Implementing extraction to OCaml could be done by connecting the $\IR$ representations with the formalisation of a suitable fragment of OCaml including type inference.
Such integration would make it possible to use the type inference algorithm to find places where coercions are necessary~\cite[Section 3.2]{letouzey04}.

\section{Acknowledgments}
This work was partially supported by the Danish Industry Foundation in the Blockchain Academy Network project.

\newpage

\lstset{language=Caml}
\begin{appendices}
\crefalias{section}{appendix}
  \section{Extracted code for the \texttt{counter} contract in Liquidity}\label{appendix:counter-liquidity}
  \begin{lstlisting}
let[@inline] fst (p : 'a * 'b) : 'a = p.(0)
let[@inline] snd (p : 'a * 'b) : 'b = p.(1)
let[@inline] addInt (i : int) (j : int) = i + j
let[@inline] subInt (i : int) (j : int) = i $-$ j
let[@inline] ltInt (i : int) (j : int) = i < j
type 'a sig_ = 'a
let exist_ a = a

type coq_msg = Coq_Inc of int | Coq_Dec of int
type coq_SimpleCallCtx = (timestamp * (address * (tez * tez)))
type storage = int
type coq_sumbool = Coq_left | Coq_right

let coq_my_bool_dec (b1 : bool) (b2 : bool) = (if b1 then fun x -> if x then Coq_left else Coq_right else fun x -> if x then Coq_right else Coq_left) b2

let coq_inc_counter (st : storage) (inc : ( (int) sig_)) =
  exist_ ((addInt st ((fun x -> x) inc)))

let coq_dec_counter (st : storage) (dec : ( (int) sig_)) =
  exist_ ((subInt st ((fun x -> x) dec)))

let coq_counter (msg : coq_msg) (st : storage) =
  match msg with
  | Coq_Inc i ->
   (match coq_my_bool_dec (ltInt 0 i) true with
    | Coq_left  -> Some ([],
      ((fun x -> x) (coq_inc_counter st (exist_ (i)))))
    | Coq_right  -> None)
  | Coq_Dec i ->
     (match coq_my_bool_dec (ltInt 0 i) true with
     | Coq_left  -> Some ([], ((fun x -> x) (coq_dec_counter st (exist_ (i)))))
     | Coq_right  -> None)

let%init storage (setup : int) =
  let inner (ctx : coq_SimpleCallCtx) (setup : int) = let ctx' = ctx in
  Some setup in
  let ctx = (Current.time (),
    (Current.sender (), (Current.amount (),Current.balance ()))) in
  match (inner ctx setup) with
  | Some v -> v
  | None -> failwith ()

let wrapper param (st : storage) =
  match coq_counter param st with
  | Some v -> v
  | None -> failwith ()

let%entry main param st = wrapper param st
  \end{lstlisting}

  \section{Extracted code for the \texttt{counter} contract in CameLIGO}\label{appendix:counter-cameligo}
  \begin{lstlisting}
[@inline] let addInt (i : int) (j : int) = i + j
[@inline] let subInt (i : int) (j : int) = i - j
[@inline] let subIntTruncated (a : int) (b : int) =
            let res = a - b in if res < 0 then 0 else res
[@inline] let multInt (i : int) (j : int) = i * j
[@inline] let divInt (i : int) (j : int) = i / j
[@inline] let leInt (i : int) (j : int) = i <= j
[@inline] let ltInt (i : int) (j : int) = i < j
[@inline] let eqInt (i : int) (j : int) = i = j

[@inline] let addTez (n : tez) (m : tez) = n + m
[@inline] let subTez (n : tez) (m : tez) = n - m
[@inline] let leTez (a : tez ) (b : tez ) = a <= b
[@inline] let ltTez (a : tez ) (b : tez ) =  a < b
[@inline] let gtbTez (a : tez ) (b : tez ) =  a > b
[@inline] let eqTez (a : tez ) (b : tez ) = a = b

[@inline] let modN (a : nat ) (b : nat ) = a mod b
[@inline] let divN (a : nat ) (b : nat ) = a / b
[@inline] let eqN (a : nat ) (b : nat ) = a = b
[@inline] let lebN (a : nat ) (b : nat ) = a <= b
[@inline] let ltbN (a : nat ) (b : nat ) = a < b

[@inline] let andb (a : bool ) (b : bool ) = a && b
[@inline] let orb (a : bool ) (b : bool ) = a || b

[@inline] let eqb_time (a1 : timestamp) (a2 : timestamp) = a1 = a2
[@inline] let leb_time (a1 : timestamp) (a2 : timestamp) = a1 <= a2
[@inline] let ltb_time (a1 : timestamp) (a2 : timestamp) = a1 < a2

[@inline] let eq_addr (a1 : address) (a2 : address) = a1 = a2
let get_contract_unit (a : address) : unit contract  =
  match (Tezos.get_contract_opt a : unit contract option) with
    Some c -> c
  | None -> (failwith ("Contract not found.") : unit contract)
type chain = {
        chain_height     : nat;
        current_slot     : nat;
        finalized_height : nat;
        account_balance  : address -> nat
      }
let dummy_chain : chain = {
        chain_height     = 0n;
        current_slot     = 0n;
        finalized_height = 0n;
        account_balance  = fun (a : address) -> 0n
      }

type coq_SimpleCallCtx = (timestamp * (address * (tez * tez)))
type storage = int
type coq_msg = Coq_Inc of (int) | Coq_Dec of (int)
type coq_sumbool = Coq_Left | Coq_Right
type 'a sig_ = 'a
let exist_ (a:int) = a

let coq_bool_dec (b1 : bool) (b2 : bool) =
  (if b1 then
    fun (x : bool) ->
      if x then Coq_Left else Coq_Right
  else fun (x : bool) ->
    if x then Coq_Right else Coq_Left) b2

let coq_Transaction_none  = ([]: (operation) list)

let coq_inc_counter (st : storage) (inc : int sig_) =
  exist_ ((addInt st ((fun (x:int sig_) -> x) inc)))

let coq_dec_counter (st : storage) (dec : int sig_) =
    exist_ ((subInt st ((fun (x:int sig_) -> x) dec)))

let coq_counter (msg : coq_msg) (st : storage) = match msg with
Coq_Inc (i) -> (match coq_bool_dec true (ltInt 0 i) with
Coq_Left  -> (Some ( (coq_Transaction_none, ((fun (x:sig_) -> x) (coq_inc_counter st (exist_ (i)))))))
  | Coq_Right  -> (None: ((operation list * storage)) option))
  | Coq_Dec (i) -> (match coq_bool_dec true (ltInt 0 i) with
Coq_Left  -> (Some ( (coq_Transaction_none, ((fun (x:sig_) -> x) (coq_dec_counter st (exist_ (i)))))))
  | Coq_Right  -> (None: ((operation list * storage)) option))

let coq_counter_wrapper (c : chain) (ctx : coq_SimpleCallCtx) (s : storage)
                        (m :  (coq_msg) option) =
  let c_ = c in
  let ctx_ = ctx in
  match m with
    Some (m0) -> (coq_counter m0 s)
  | None  -> (None: ((operation list * storage)) option)

let init (setup : int) : storage =
  let inner (ctx : coq_SimpleCallCtx) (setup : int) : (storage) option =
    let ctx_ = ctx in
    Some (setup) in
  let ctx = (Tezos.now,
    (Tezos.sender,
    (Tezos.amount,
    Tezos.balance))) in
  match (inner ctx setup) with
    Some v -> v
  | None -> (failwith (""): storage)

type init_args_ty = int
let init_wrapper (args : init_args_ty) = init args

type return = (operation) list * (storage option)
type parameter_wrapper = Init of init_args_ty | Call of coq_msg option

let wrapper (param, st : parameter_wrapper * (storage) option) : return =
  match param with
    Init init_args -> (([]: operation list), Some (init init_args))
  | Call p -> (
    match st with
      Some st -> (match (coq_counter_wrapper dummy_chain (Tezos.now,
    (Tezos.sender,
    (Tezos.amount,
    Tezos.balance))) st p) with
                    Some v -> (v.0, Some v.1)
                  | None -> (failwith ("") : return))
    | None -> (failwith ("cannot call this endpoint before Init has been called"): return))
let main (action, st : parameter_wrapper * storage option) : return = wrapper (action, st)
  \end{lstlisting}
\end{appendices}

\bibliographystyle{agsm}
\bibliography{paper}
\label{lastpage01}

\end{document}